\documentclass[fleqn,usenatbib]{mnras}

\usepackage{newtxtext,newtxmath}

\usepackage[T1]{fontenc}
\usepackage{ae,aecompl}

\usepackage{graphicx}	
\usepackage{amsmath}	
\usepackage{braket}

\title[]{The generation and transformation of polarisation signals in molecular lines through collective anisotropic resonant scattering}

\author[M. Houde et al.]{M. Houde,$^{1}$\thanks{E-mail: mhoude2@uwo.ca} B. Lankhaar,$^{2}$ F. Rajabi,$^{3,1}$ M. A. Chamma$^{1}$  
\\
$^{1}$Department of Physics and Astronomy, The University of Western Ontario, 1151 Richmond Street, London, Ontario N6A 3K7, Canada\\
$^{2}$Department of Space, Earth and Environment, Chalmers University of Technology, Onsala Space Observatory, 439 92 Onsala, Sweden \\
$^{3}$Perimeter Institute for Theoretical Physics, 31 Caroline Street N, Waterloo, Ontario N25 2YL, Canada
}

\date{Accepted 2021 December 29. Received 2021 December 17; in original form 2021 September 20}

\pubyear{2020}

\begin{document}
\label{firstpage}
\pagerange{\pageref{firstpage}--\pageref{lastpage}}
\maketitle

\begin{abstract}
We discuss the existence of elliptical polarisation in rotational spectral lines of $\mathrm{CO}$ and other molecules within the context of the Anisotropic Resonant Scattering (ARS) model. We show that the effect of ARS on the radiation field can lead to not only the previously predicted transformation of background linear polarisation into circular polarisation (i.e., Faraday conversion), but also the occurrence of Faraday rotation and the generation of elliptically polarised signals in an otherwise initially unpolarised radiation field. This is due to a collective behaviour between the large number of molecules acting as a diffraction ensemble that strongly favours forward scattering over any other mode. Our application to astronomical data demonstrates the dependency of the Stokes parameters on the strength and orientation of the ambient magnetic field, and suggests that ARS will manifest itself for a wide range of molecular species and transitions.
\end{abstract}




\section{Introduction}\label{sec:Introduction}
Given the challenges involved in acquiring direct measurements of magnetic field strengths through the Zeeman effect over a broad range of sources \citep{Crutcher2012}, studies aimed at characterizing the importance and the role of magnetic fields in the star formation process, or in other environments of the interstellar medium, have long relied on indirect statistical methods \citep{Davis1951,Chandrasekhar1953}. Accordingly, tools for the analysis of linear polarisation maps from dust emission have been developed for pursuing these goals and have provided much of the observational input in the recent past (see, e.g., \citealt{Lazarian2007,Hildebrand2009,Houde2009,Soler2013,Houde2016}). 

The prospect and subsequent realisation of dust polarization maps at very high spatial resolutions with ALMA promised to push these studies well within the domain of circumstellar disks and, furthermore, help elucidate the process of planet formation. Although some definite progress has been made through corresponding observations of protostellar systems (see, e.g., \citealt{Maury2018,Hull2019}), the use of dust polarisation maps at the finest scales has been limited by the realisation that self-scattering of dust emission by larger grains in circumstellar disks can remove the correlation between the relative orientations of ambient magnetic fields and measured linear polarisation (pseudo-)vectors \citep{Kataoka2015,Yang2016}. Similarly, strong radiative alignment torques can set dust grains in a precession about the radiation anisotropy vector rather than that of the magnetic field (see \citealt{Pattle2021} and further references therein). A natural solution to this problem would thus be to rely on linear polarization maps from molecular spectral lines obtained in the same environments. It has long been known that several species/transitions show linear polarization signals in their spectra \citep{Glenn1997}. These are expected to be well correlated with the orientation of the ambient magnetic field responsible for the underlying molecular alignment\footnote{The expression ``molecular alignment'' is used in this paper to mean that the symmetry axis of a molecule is aligned with the ambient magnetic field. In general, it would be more precise to reserve the term ``alignment'' for situations when (almost) degenerate magnetic sub-levels have unequal populations, so that the angular momentum vector, $\boldsymbol{j}$, has a preferred direction (with respect to the symmetry axis, $\boldsymbol{B}$). One speaks about ``orientation'' when $\left<\hat{\boldsymbol{j}}\cdot \hat{\boldsymbol{B}}\right> \neq 0$, while alignment is said to be present when $\left<\left|\hat{\boldsymbol{j}}\cdot \hat{\boldsymbol{B}}\right|^2\right> \neq 2/3$ \citep{Landi2004,Blum2012}}, presumably through the so-called Goldreich-Kylafis effect \citep{Goldreich1981}. 

However, the initial discovery by \citet{Houde2013} of circular polarisation signals in the $\mathrm{CO}\left(J=2\rightarrow1\right)$ transition (weakly sensitive to the Zeeman effect) at the peak intensity position in Orion KL, as well as the subsequent detections in the same line (and in $\mathrm{CO}\left(J=1\rightarrow0\right)$) in SNR IC 443 by \citet{Hezareh2013} have unveiled a more complicated picture for the nature of polarisation found in molecular spectra. More precisely, most models developed that either predicted or attempted to explain the existence of such circular polarisation in molecular lines implied a conversion of background linear polarisation into circular polarisation \citep{Deguchi1985,Wiebe1998,Houde2013}, which results in a rotation of the linear polarisation vectors and a loss of their correlation with the ambient magnetic field's orientation. This rotation in the orientation of linear polarisation signals was confirmed with the SNR IC 443 observations of \citet{Hezareh2013}, who noted a lack of correlation in the corresponding orientations between maps of $\mathrm{CO}\left(J=1\rightarrow0\right)$, $\mathrm{CO}\left(J=2\rightarrow1\right)$ and dust polarisation at $870\,\micron$. Most interestingly, their reinsertion of the circular polarisation also detected in the two CO transitions into the corresponding linear polarisation signals (by reversing the aforementioned polarisation conversion process) reestablished a near perfect correlation between the orientation of linear polarisation vectors in the CO maps, as well as with the dust polarisation map. 

The unavoidable consequence stemming from the work of \citet{Hezareh2013} is that one should not blindly use linear polarisation measurements from molecular lines when applying statistical techniques initially developed for dust polarisation maps without properly accounting for the presence of circular polarisation. At the very least, one should first verify if circular polarisation signals are present before attempting such analyses (see, e.g., \citealt{Ching2016}). This is to be kept in mind as more measurements of circular polarisation in molecular lines from ALMA become available \citep{Vlemmings2019}, while a recent search through the SMA archive suggests that the presence of circular polarisation signals may be relatively common for several molecular species/transitions \citep{Chamma2018}. 

A further implication arising from these considerations is the necessity of developing a better understanding of the physical process(es) responsible for the existence of these polarisation signals in molecular lines. Accordingly, in this paper we revisit the Anisotropic Resonant Scattering (ARS) model of \citet{Houde2013} developed to explain the initial discovery of circular polarisation in $\mathrm{CO}\left(J=2\rightarrow1\right)$ in Orion KL with the goal of setting it on a more rigorous theoretical footing (this model was also considered in \citealt{Hezareh2013} and \citealt{Chamma2018}, and applied to the SiO maser observations of \citealt{Cotton2011} by \citealt{Houde2014}). We start in Sec. \ref{sec:theory} by considering an analysis for the time evolution of an initial quantum mechanical state describing a molecular ensemble and an accompanying radiation field, and investigate the radiation-matter interaction due to absorption/emission and resonant scattering. In Sec. \ref{sec:ARS} we consider the effect of ARS on the state of the radiation field, and we show how it can lead to not only the transformation of background linear polarisation into circular polarisation (i.e., Faraday conversion) predicted by \citet{Houde2013}, but also the occurrence of Faraday rotation and the generation of elliptically polarised signals in an otherwise initially unpolarised radiation field. We further calculate the dependency of the Stokes parameters on the strength and orientation of the ambient magnetic field responsible for the alignment of the molecular population. In Sec. \ref{sec:application} we apply our analysis to astronomical data originally presented in \citet{Houde2013}, while we summarize our results in Sec. \ref{sec:summary}. Appendices detailing some calculations can be found at the end. 

\section{Analysis}\label{sec:theory}

Before discussing the generation and transformation of polarisation signals through ARS, which involves the differences in the scattering amplitudes between radiation modes of perpendicular polarisation states, we must first consider the time evolution of a single quantum mechanical state. The system we have in mind is one composed of a large group of $N\gg 1$ molecules (e.g., CO) occupying a volume of linear size $L\gg \lambda$, with $\lambda$ the (average) wavelength of radiation, being irradiated by an electromagnetic field incident from a background location and propagating in the general direction of an observer far away in the foreground. On its way to the observer the radiation field interacts with the molecular population, which in the process alters its polarisation state. We aim to characterise the polarisation state of the emerging radiation as detected by the observer.  

We therefore start with an initial and normalized compound molecular/radiation state
\begin{equation}
\Ket{\psi_0}=\sum_{p}c_{p}\Ket{u_{p}},\label{eq:|psi>_ars}
\end{equation}
\noindent where the basis states are of the type $\Ket{u_{p}}\equiv\Ket{m}\Ket{n_{p}}$, with $\Ket{m}$ a global molecular state (see below and Appendix \ref{app:global_state}) and $\Ket{n_{p}}$ the incident radiation state with a large number of photons occupying a set of radiation modes characterized by corresponding wave vectors and one of two perpendicular polarisation states. For linear polarisation these states will be denoted by $\Vert$ and $\bot$ for orientations respectively parallel and perpendicular to the direction of the plane of the sky component of the local magnetic field responsible for molecular alignment. A pair of orthogonal circular, denoted by the ``$R$'' and ``$L$'' subscripts, as well as a pair of orthogonal elliptical polarisation states will also be used. Although there are in principle several energy levels available to any molecule, we are interested in radiative transitions between sets of potentially degenerate lower $\left\{ \Ket{g_j}\right\}$ and excited $\left\{ \Ket{e_k}\right\}$ states of energies $E_{g}$ and $E_{e}$, respectively. 

Our analysis will focus on transitions in and out of states that have $N_g$ molecules in a lower state $\Ket{g_j}$ and $N_e$ in an excited state $\Ket{e_k}$. We therefore define the quantum number 
\begin{equation}
    m = \frac{N_e-N_g}{2},\label{eq:m}
\end{equation}
and label by $\Ket{m}$ the aforementioned ket for the group of corresponding degenerate global internal molecular states. The transitions entering our analysis are thus of the type $\Ket{m}\leftrightarrow\Ket{m}$ for scattering and $\Ket{m}\leftrightarrow\Ket{m\pm1}$ for the absorption/emission of a photon, where there are respectively $N_e\pm1$ molecules in excited and $N_g\mp1$ in lower states for $\Ket{m\pm1}$.  

We define the radiative state $\Ket{n_{p}}$ as containing $n_{\gamma}$ photons, and could be represented as
\begin{equation}
\Ket{n_{p}} = \Ket{n_{1},n_{2},\ldots,n_{w},\ldots}\label{eq:|np>}
\end{equation}
\noindent with $n_{\gamma}=n_{1}+n_{2}+\ldots+n_{w}+\ldots$ and where $\Ket{n_{1}}$ is the state corresponding to mode 1 made of $n_{1}$ photons characterized by a wave vector $\mathbf{k}_{1}$ and a given polarization state (e.g., $\Vert$, $\bot$, $R$, $L$, ...), etc.

Our analysis will consist of tracking the temporal evolution of the system using the evolution operator
\begin{equation}
    \hat{U}\left(t,t_0\right) = e^{-i\hat{H}\left(t-t_0\right)/\hbar}\label{eq:U}
\end{equation}
\noindent such that the state of the system at time $t$ is \citep{Cohen-Tannoudji2018a}
\begin{equation}
    \Ket{\psi\left(t\right)} = \hat{U}\left(t,t_0\right)\Ket{\psi_0}\label{eq:|psi(t)>}
\end{equation}
with $\Ket{\psi_0}$ the state of the system at time $t_0$. In equation (\ref{eq:U}) the Hamiltonian is given by
\begin{equation}
    \hat{H} = \hat{H}_0+\hat{W},\label{eq:H}
\end{equation}
\noindent where $\hat{H}_0$ consists of the sum of the independent (non-interacting) molecular and radiation Hamiltonians, as well as the Zeeman Hamiltonian responsible for the corresponding splitting that lifts the degeneracy between the $\pi$ and $\sigma_\pm$ spectral lines to be considered later. The interaction Hamiltonian $\hat{W}$ arises from the electric dipole interaction through
\begin{align}
\hat{W} = & -i\sum_{q}\mathcal{E}_{q}\sum_{\alpha=1}^{N}\hat{\mathbf{d}}_{\alpha}\cdot\left[\boldsymbol{\epsilon}_{q}\hat{a}_{q}e^{i\mathbf{k}_{q}\cdot\mathbf{r}_{\alpha}}-\boldsymbol{\epsilon}_{q}^*\hat{a}_{q}^{\dagger}e^{-i\mathbf{k}_{q}\cdot\mathbf{r}_{\alpha}}\right],\label{eq:W}
\end{align}
\noindent where $\mathcal{E}_{q}=\sqrt{\hbar\omega_{q}/2\epsilon_{0}L^{3}}$ is the one-photon electric field, with $L^{3}$ the quantization volume (determined by the volume containing the $N=N_g+N_e$ molecules potentially interacting with the radiation), $\hat{\mathbf{d}}_{\alpha}$ the molecular electric  dipole moment of molecule $\alpha$ and $\hat{a}_{q}^{\dagger}$ ($\hat{a}_{q}$) the photon creation (annihilation) operator for the radiation mode $q$. The unit vector $\mathbf{\boldsymbol{\epsilon}}_{q}$ is for the radiation polarisation states,  $\mathbf{r}_{\alpha}$ is the position of molecule $\alpha$, while the frequencies $\omega_{\alpha}$ and $\omega_{q}$ are associated with the molecular transition and the radiation, respectively. Accordingly, the summations on $q$ and $\alpha$ are for radiation (wave vector and polarization) and molecular states, respectively \citep{Grynberg2010}.

Our analysis will be facilitated by using the following (exact) expansion for the evolution operator \citep{Cohen-Tannoudji2017}
\begin{align}
    \hat{U}\left(t,t_0\right) & = \hat{U}_0\left(t,t_0\right)+\frac{1}{i\hbar}\int_{t_0}^t dt^\prime\hat{U}_0\left(t,t^\prime\right)\hat{W}\hat{U}_0\left(t^\prime,t_0\right)\nonumber\\
    & -\frac{1}{\hbar^2}\int_{t_0}^t dt^\prime\int_{t_0}^{t^\prime}dt^{\prime\prime}\hat{U}_0\left(t,t^\prime\right)\hat{W}\hat{U}\left(t^\prime,t^{\prime\prime}\right)\hat{W}\hat{U}_0\left(t^{\prime\prime},t_0\right),\label{eq:Uexp}
\end{align}
\noindent with
\begin{equation}
     \hat{U}_0\left(t,t_0\right) = e^{-i\hat{H}_0\left(t-t_0\right)/\hbar}.\label{eq:U0}
\end{equation}
\noindent We note that the states $\Ket{u_p}=\Ket{m}\Ket{n_p}$, etc., are eigenstates of $\hat{H}_0$ with eigenvalues $E_p^0$ (see below). 

The state of the system at time $t$ is given by
\begin{equation}
    \Ket{\psi\left(t\right)} = \sum_{p}\left[c_{p}^{\left(0\right)}\left(t\right)+c_{p}^{\left(1\right)}\left(t\right)+c_{p}^{\left(2\right)}\left(t\right)\right]\Ket{u_{p}}\label{eq:|psi(t)_exp}
\end{equation}
\noindent with, from equation (\ref{eq:Uexp}) (setting $t_0=0$ for convenience),
\begin{align}
    c_{p}^{\left(0\right)}\left(t\right) & = \braket{u_p|\hat{U}_0\left(t,0\right)|\psi_0}\nonumber\\
    & = c_{p}e^{-iE_{p}^{0}t/\hbar}\label{eq:c0(t)}\\
    c_{p}^{\left(1\right)}\left(t\right) & = \frac{1}{i\hbar}\int_0^t dt^\prime\braket{u_p|\hat{U}_0\left(t,t^\prime\right)\hat{W}\hat{U}_0\left(t^\prime,0\right)|\psi_0}\nonumber\\
    & = -\frac{it}{\hbar}e^{-iE_{p}^{0}t/\hbar}\sum_{n}c_n W_{pn}e^{i\frac{1}{2}\omega_{pn}t}\mathrm{sinc}\left(\frac{1}{2}\omega_{pn}t\right)\label{eq:c1(t)}\\
    c_{p}^{\left(2\right)}\left(t\right) & = -\frac{1}{\hbar^2}\int_0^t dt^\prime\nonumber\\
    & \quad\times\int_0^{t^\prime}dt^{\prime\prime}\braket{u_p|\hat{U}_0\left(t,t^\prime\right)\hat{W}\hat{U}\left(t^\prime,t^{\prime\prime}\right)\hat{W}\hat{U}_0\left(t^{\prime\prime},0\right)|\psi_0}\nonumber\\
    & = -\frac{1}{\hbar^2}e^{-iE_{p}^{0}t/\hbar}\sum_{n}\sum_{j,k}c_n W_{pj}W_{kn}\int_0^t dt^\prime e^{iE_{p}^{0}t^\prime/\hbar}\nonumber\\
    & \quad\times\int_0^{t^\prime}dt^{\prime\prime}e^{-iE_{n}^{0}t^{\prime\prime}/\hbar}\braket{u_j|\hat{U}\left(t^\prime,t^{\prime\prime}\right)|u_k},\label{eq:c2(t)}
\end{align}
\noindent where $\omega_{pn}=\left(E_{p}^{0}-E_{n}^{0}\right)/\hbar$, with $E_{p}^{0}$ the unperturbed energy level of state $\Ket{u_{p}}$, and the matrix elements $W_{pn}=\Braket{u_p|\hat{W}|u_n}$ \citep{Cohen-Tannoudji2017,Cohen-Tannoudji2018a}. The summations on $n$, $j$ and $k$ are for complete states, i.e., composed of molecular and radiation components as defined above.

We now turn to the evaluation of the probability amplitudes and probabilities for the absorption/emission and resonant scattering processes by solving equations (\ref{eq:c1(t)}) and (\ref{eq:c2(t)}), respectively, starting with the former. 

\subsection{Absorption/emission}\label{sec:absorption}

Calculations for the probability amplitude for absorption/emission pertain to the first-order term in $\hat{W}$ given in equation (\ref{eq:c1(t)}). For the initial state defined in equation (\ref{eq:|psi>_ars}), the corresponding first-order state is (i.e., including the zeroth-order term of equation (\ref{eq:c0(t)}))
\begin{align}
\Ket{\psi^{\left(1\right)}\left(t\right)} = & \sum_{p}e^{-iE_{p}^{0}t/\hbar}\left[c_{p}-\frac{it}{\hbar}\sum_{n}c_{n}W_{pn}\right.\nonumber \\
 & \qquad\qquad\quad\left.\rule{0mm}{6mm}\times e^{i\frac{1}{2}\omega_{pn}t}\mathrm{sinc}\left(\frac{1}{2}\omega_{pn}t\right)\right]\Ket{u_{p}},\label{eq:|psi'>_abs}
\end{align}
\noindent where, as stated before, we focus on transitions between states containing $N_g$ (i.e., $\Ket{m}$) and $N_g\mp1$ (i.e., $\Ket{m\pm1}$) molecules in lower states. More precisely, we consider a change in states that involves a transition between two states $\Ket{g_j}$ and $\Ket{e_k}$; the respective degeneracy are denoted by $g_g$ and $g_e$. As shown in Appendix \ref{app:absorption}, considering a specific state $\Ket{u_p}=\Ket{m}\Ket{n_p}$ while distinguishing between cases where transitions happen into and out of state $\Ket{u_p}$ allows us to rewrite equation (\ref{eq:|psi'>_abs}) as  
\begin{align}
    \Ket{\psi^{\left(1\right)}\left(t\right)} \simeq & \,e^{-iE_{p}^{0}t/\hbar}\left\{\rule{0mm}{7mm}c_{p}+\frac{t}{\hbar}\sum_{r}\,\mathcal{E}_{r}\right.\nonumber \\
    & \hspace{-14.5mm} \times\left[\sqrt{n_{r}+1}\left(\mathbf{\boldsymbol{\epsilon}}_{d}^*\cdot\boldsymbol{\epsilon}_{r}^*
\right)d^*\sum_{\beta=1}^{N_{e}+1}c_{r\beta}e^{-i\mathbf{k}_{r}\cdot\mathbf{r}_{\beta}}e^{-i\frac{1}{2}\omega_{\beta r}t}\right.\nonumber\\
    & \hspace{-14mm} \left.-\sqrt{n_{r}}\left(\mathbf{\boldsymbol{\epsilon}}_{d}\cdot\boldsymbol{\epsilon}_{r}
\right)d\sum_{\beta=1}^{N_{g}+1}c_{r\beta}e^{i\mathbf{k}_{r}\cdot\mathbf{r}_{\beta}}e^{i\frac{1}{2}\omega_{\beta r}t}\right]\left.\mathrm{sinc}\left(\frac{1}{2}\omega_{\beta r}t\right)\rule{0mm}{7mm}\right\}\Ket{u_{p}}\nonumber \\
    & \hspace{-14.5mm} -c_{p}\frac{t}{\hbar}\sum_{r}\,\mathcal{E}_{r}\left[\sqrt{n_{r}+1}\left(\mathbf{\boldsymbol{\epsilon}}_{d}\cdot\boldsymbol{\epsilon}_{r}\right)d\sum_{\beta=1}^{N_{g}}e^{i\mathbf{k}_{r}\cdot\mathbf{r}_{\beta}}e^{i\frac{1}{2}\omega_{\beta r}t}\Ket{m+1}\right.\nonumber \\
    & \hspace{-14mm} \left.-\sqrt{n_{r}}\left(\mathbf{\boldsymbol{\epsilon}}_{d}^*\cdot\boldsymbol{\epsilon}_{r}^*\right)d^*\sum_{\beta=1}^{N_{e}}e^{-i\mathbf{k}_{r}\cdot\mathbf{r}_{\beta}}e^{-i\frac{1}{2}\omega_{\beta r}t}\Ket{m-1}\right]\nonumber\\
    & \hspace{-13mm} \times e^{-iE_{n}^{0}t/\hbar}\mathrm{sinc}\left(\frac{1}{2}\omega_{\beta r}t\right)\Ket{n_{r}},\label{eq:|psi'>_abs-1}
\end{align}
with
\begin{equation}
    \omega_{\beta r} \simeq \omega_{\beta}-\omega_{r}\left(1-\frac{\mathbf{v}_\beta\cdot\mathbf{e}_{r}}{c}\right),\label{eq:omega_beta-r_main}
\end{equation}
\noindent where $\mathbf{v}_\beta$ is the velocity of molecule $\beta$ and $\mathbf{e}_r=\mathbf{k}_r/k_r$. In equation (\ref{eq:|psi'>_abs-1}) all molecules share the same dipole moment $d$ and corresponding unit vector $\mathbf{\boldsymbol{\epsilon}}_{d}$ (both defined for an upward transition, i.e., from a lower $\Ket{g_j}$ to an excited $\Ket{e_k}$ state; $d^*$ and $\mathbf{\boldsymbol{\epsilon}}_{d}^*$ are therefore for the reverse transition) and we substituted $\sum_{n}c_{n}\rightarrow \sum_{r}\sum_ {\beta}c_{r\beta}$ ($\beta$ specifies the molecule and $r$ the radiation mode; see Appendix \ref{app:absorption}). The first three lines are for transitions from states $\Ket{u_n}\equiv\Ket{m\pm1}\Ket{n_r}$ into $\Ket{u_p}=\Ket{m}\Ket{n_p}$ while the last three concern the opposite transitions, and we omitted terms not involving the state $\Ket{u_p}$. For example, this equation reveals that all states which initially share the same radiation modes as $\Ket{u_{p}}$ but differ in only one of them with one less photon, and also share all of the same molecular states except for having one more of these molecules in an excited state can contribute to the probability amplitude of $\Ket{u_{p}}$ through the emission of a photon by that molecule into that radiation mode (the first term within the first pair of square brackets). In the same manner, $\Ket{u_{p}}$ can lose a photon to the other states through absorption (the first term within the second pair of square brackets).

It is interesting to note that the presence of a large number of randomly located molecules in the volume of interaction has an important consequence on the expected value of the probability amplitude. To better see this we make the simplification where all states are equally probable, i.e., $c_{r\beta}=c_{p}=c_0$ a constant. This is equivalent to stating that all combinations for $r$ and $\beta$ in the double summations $\sum_{r}\sum_{\beta}$, are realized and, therefore, $e^{\pm i\frac{1}{2}\omega_{\beta r}t}\mathrm{sinc}\left(\frac{1}{2}\omega_{\beta r}t\right)\simeq1$. We could then rewrite for the term corresponding to the absorption of a photon from a state $\Ket{m}$ to a state $\Ket{m+1}$, for example, 
\begin{align}
d\sum_r\sqrt{n_{r}+1}\mathcal{E}_{r} & \left(\mathbf{\boldsymbol{\epsilon}}_{d}\cdot\boldsymbol{\epsilon}_{r}\right)\sum_{\beta=1}^{N_g}e^{i\mathbf{k}_{r}\cdot\mathbf{r}_{\beta}}e^{-i\frac{1}{2}\omega_{\beta r}t}\mathrm{sinc}\left(\frac{1}{2}\omega_{\beta r}t\right)\nonumber \\ 
& = d\sum_r\sqrt{n_{r}+1}\mathcal{E}_{r}\left(\mathbf{\boldsymbol{\epsilon}}_{d}\cdot\boldsymbol{\epsilon}_{r}\right)\sum_{\beta=1}^{N_g}e^{i\mathbf{k}_{r}\cdot\mathbf{r}_{\beta}}.\label{eq:sum_n-1}
\end{align}
\noindent We recognize on the right-hand side of this relation a summation on all molecules, which we assume to be randomly spread within the volume. We now consider the expected value of the summation
\begin{equation}
\left\langle \sum_{\beta=1}^{N_{g}}e^{i\mathbf{k}_{r}\cdot\mathbf{r}_{\beta}}\right\rangle  = \sum_{\beta=1}^{N_{g}}\left\langle e^{i\mathbf{k}_{r}\cdot\mathbf{r}_{\beta}}\right\rangle,\label{eq:sum_beta-random-1}
\end{equation}
\noindent where each of these averages are over random phase terms such that
\begin{align}
\left\langle e^{i\mathbf{k}_{r}\cdot\mathbf{r}_{\beta}}\right\rangle & = \frac{1}{2\pi}\int_{0}^{2\pi}e^{i\varphi}d\varphi\nonumber \\
 & = 0.\label{eq:sum_beta-random}
\end{align}
\noindent This implies that, on average, the corresponding probability amplitude will cancel out to zero. The same applies for all processes in equation (\ref{eq:|psi'>_abs-1}). This, of course, does not mean that the first order processes of emission and absorption are unimportant since, as we will now see, fluctuations about this average will be responsible for a probability scaling with the number of molecules. Rather, equation (\ref{eq:sum_beta-random}) implies that for a large ensemble of randomly positioned molecules emission and absorption events are non-coherent. This is radically different from what will be found for the forward resonant scattering process, which is coherent in nature, and leads to a non-vanishing probability amplitude proportional to the number of molecules.

\begin{figure}
\begin{center}\includegraphics[scale=0.58]{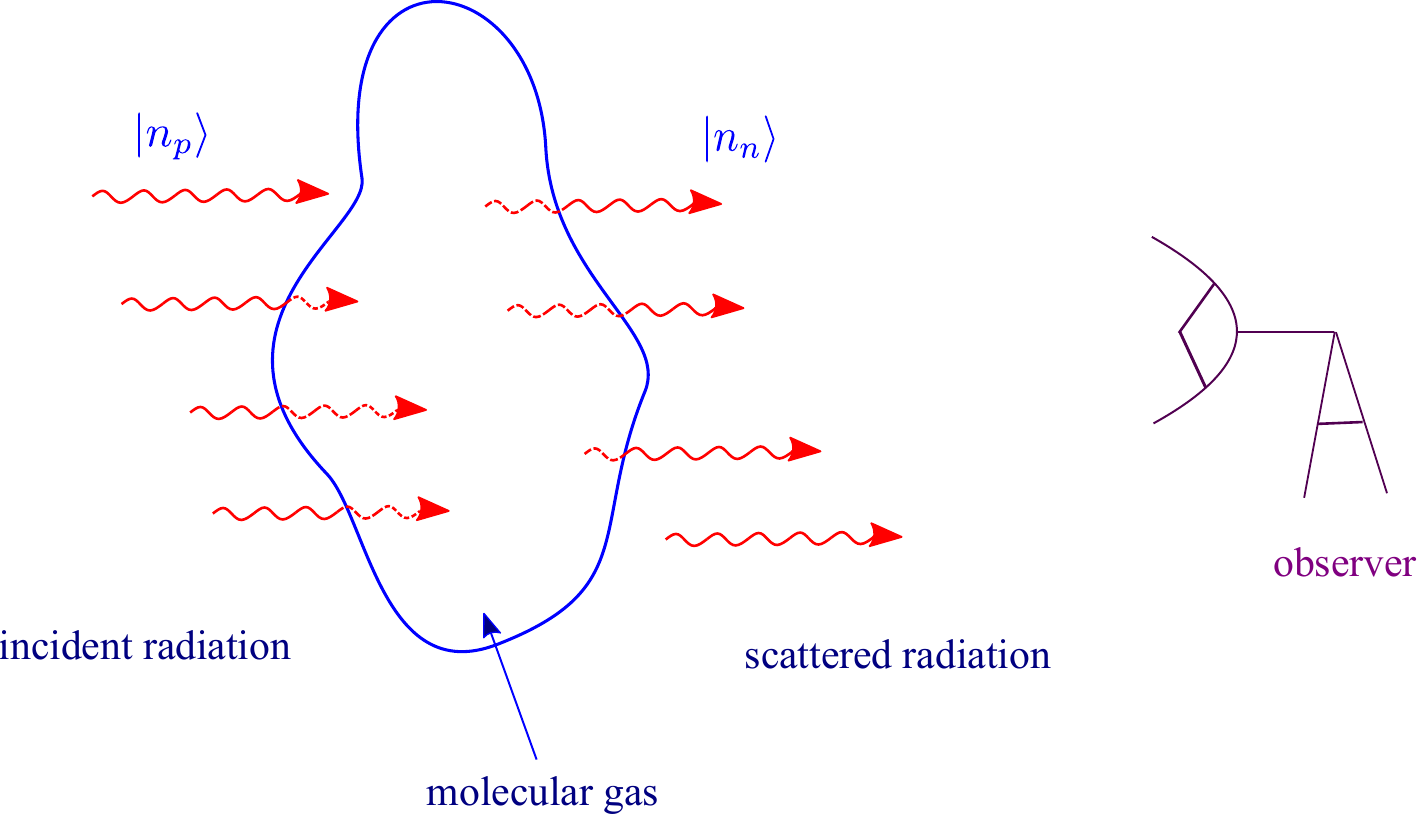}\end{center}

\caption{\label{fig:ARS}Representation of incident radiation $\Ket{n_{p}}$ basis states scattered into states $\Ket{n_{n}}$ by a molecular gas and detected by the observer. The radiation modes in each state points toward the observer and, at a given frequency $\omega_{q}$, all such modes (mode $q$) contribute to the signal detected by an observer. Our analysis shows that $\Ket{n_{n}}=\Ket{n_{p}}$.}
\end{figure}

To calculate the probability associated with the different absorption and emission processes present in equation (\ref{eq:|psi'>_abs-1}) we again assume all states to be equally probable (i.e., $c_{r\beta}=c_{p}=c_0$) while we focus on radiation modes pointing in the direction toward the observer, as shown in Figure \ref{fig:ARS}, since only them will factor in the calculations. Squaring the norm of the probability amplitude brings in summations of the type
\begin{align}
    \left\langle \sum_{\alpha,\beta=1}^{N_{g}}e^{i\mathbf{k}_{q}\cdot\left(\mathbf{r}_{\alpha}-\mathbf{r}_{\beta}\right)}\right\rangle & = \sum_{\alpha=1}^{N_{g}}+ \sum_{\alpha\neq\beta}^{N_g}\left\langle e^{i\mathbf{k}_{q}\cdot\left(\mathbf{r}_{\alpha}-\mathbf{r}_{\beta}\right)}\right\rangle \nonumber \\
    & = N_{g}\int h\left(\omega_{\alpha}\right)d\omega_{\alpha},\label{eq:sum_alpha-beta}
\end{align}
\noindent which do not vanish on account of the first term on the right-hand side of the first line. In this equation $h\left(\omega_{\alpha}\right)$ is the normalized probability distribution function of the molecular Bohr frequency (velocity).

Detailed calculations are given in Appendix \ref{app:absorption} for the evaluation of the probabilities and transition rates with the processes contained in equation (\ref{eq:|psi'>_abs-1}). Most relevant to our analysis are the total decay rate associated to $\Ket{u_p}=\Ket{m}\Ket{n_p}$ 
\begin{align}
    & \Gamma_{p}^{-}\left(\omega_{q}\right) = \Gamma_{\mathrm{abs}}^m\left(\omega_{q}\right)+\Gamma_{\mathrm{em}}^m\left(\omega_{q}\right)\label{eq:Gamma-a}
\end{align}
\noindent for the absorption and emission rates out of the molecular state $\Ket{m}$, respectively, as well as its total enhancement rate
\begin{align}
    & \Gamma_{p}^{+}\left(\omega_{q}\right) = \Gamma_{\mathrm{em}}^{m+1}\left(\omega_{q}\right)+\Gamma_{\mathrm{abs}}^{m-1}\left(\omega_{q}\right)\label{eq:Gamma+a}
\end{align}
\noindent for emission from $\Ket{m+1}$ and absorption from $\Ket{m-1}$, where
\begin{align}
    & \Gamma_{\mathrm{abs}}^j\left(\omega_{q}\right) \simeq \left(\frac{N}{2}-j\right)\frac{3\pi^{2}c^{3}}{\hbar\omega_{0}^{3}}A_{eg}u\left(\omega_{q}\right)h\left(\omega_{q}\right)\left|\mathbf{\boldsymbol{\epsilon}}_{d}\cdot\boldsymbol{\epsilon}_{q}\right|^2\label{eq:Gamma_abs_a}\\
    & \Gamma_{\mathrm{em}}^j\left(\omega_{q}\right)\simeq \frac{g_g}{g_e}\left(\frac{N}{2}+j\right)\frac{3\pi^{2}c^{3}}{\hbar\omega_{0}^{3}}A_{eg}u\left(\omega_{q}\right)h\left(\omega_{q}\right)\left|\mathbf{\boldsymbol{\epsilon}}_{d}\cdot\boldsymbol{\epsilon}_{q}\right|^2\label{eq:Gamma_se_a}
\end{align}
\noindent and, once again, $N=N_g+N_e$ and $j=m, m\pm1$ where $m$ is defined in equation (\ref{eq:m}). In equations (\ref{eq:Gamma_abs_a})-(\ref{eq:Gamma_se_a}) we introduced the energy density $u\left(\omega_{q}\right)$ at $\omega_q$ and the Einstein spontaneous emission coefficient $A_{eg}$ for the $\Ket{e_k}\rightarrow\Ket{g_j}$ transition at the systemic velocity (i.e., at the centre frequency $\omega_0$ of the distribution $h\left(\omega_\alpha\right)$)
\begin{align}
    u\left(\omega_{q}\right) & = \left(\frac{n_{q}\hbar\omega_{q}}{L^{3}}\right)\nonumber \\
    & = 2\epsilon_{0}n_{q}\mathcal{E}_{q}^{2}\label{eq:u(omega_p)}\\
    A_{eg} & = \frac{\omega_{0}^{3}\left|d\right|^{2}}{3\pi\epsilon_{0}\hbar c^{3}}.\label{eq:Aeg}
\end{align}

To consider the evolution of the radiation field on its own, we can set the molecular state to be $\Ket{m}$ (i.e., set the number of molecules in the lower and upper states fixed at $N_g$ and $N_e$, respectively) and compare (i.e., subtract) the stimulated emission rate $\Gamma_{\mathrm{em}}^m\left(\omega_{q}\right)$ from the absorption rate $\Gamma_{\mathrm{abs}}^m\left(\omega_{q}\right)$. Dividing these by $c$ and setting $\left|\mathbf{\boldsymbol{\epsilon}}_{d}\cdot\boldsymbol{\epsilon}_{q}\right|^2=1$ yields the absorption coefficient for a lone photon at frequency $\omega_{q}$
\begin{equation}
    \alpha\left(\omega_{q}\right)\simeq n_{g}\left(1-\frac{g_gn_e}{g_en_g}\right)\frac{3\pi^{2}c^{2}}{\omega_{0}^{2}}A_{eg}h\left(\omega_{q}\right),\label{eq:abs_coef}
\end{equation}
\noindent with $n_{g}$ and $n_e$ the density of molecules in the lower and upper states, respectively. For example, at thermodynamic equilibrium
\begin{equation}
    n_g = n\frac{g_g e^{-E_g/kT}}{Q\left(T\right)}\label{eq:Boltzmann}
\end{equation}
\noindent with $n$ the molecular density and $Q\left(T\right)$ the partition function at temperature $T$. Equation (\ref{eq:abs_coef}) is similar to what can be found elsewhere in the literature (see \citealt{Rybicki1979} for a semi-classical derivation, for example).

\subsection{Resonant scattering}\label{scattering}

For the resonant scattering process we must consider equation (\ref{eq:c2(t)}) for the second-order term in $\hat{W}$, which we write here again for convenience
\begin{align}
    c_{p}^{\left(2\right)}\left(t\right) & = -\frac{1}{\hbar^2}e^{-iE_{p}^{0}t/\hbar}\sum_{n}\sum_{j,k}c_n W_{pj}W_{kn}\int_0^t dt^\prime e^{iE_{p}^{0}t^\prime/\hbar}\nonumber\\
    & \quad\times\int_0^{t^\prime}dt^{\prime\prime}e^{-iE_{n}^{0}t^{\prime\prime}/\hbar}\braket{u_j|\hat{U}\left(t^\prime,t^{\prime\prime}\right)|u_k}.\label{eq:c2(t)_again}
\end{align}

For the matrix elements $W_{pj}$ and $W_{kn}$ on the right-hand side of equation (\ref{eq:c2(t)_again}) the subscripts $p$ and $n$ are for a molecular state containing, as before, $N_{g}$ molecules in the $\Ket{g_j}$ lower states and a predetermined number of photons, say, $n_{\gamma}$ spread among the radiation modes. More precisely, we set $\Ket{u_p}=\Ket{m}\Ket{n_p}$  and $\Ket{u_n}=\Ket{m^\prime}\Ket{n_n}$, where both radiation states contain $n_\gamma$ photons while the molecular states $\Ket{m}$ and $\Ket{m^\prime}$ are initially assumed different (i.e., both states contain $N_g$ molecules in a lower state but with potentially different arrangements). The intermediate (virtual) states $\Ket{u_j}$ and $\Ket{u_k}$ contain $N_{g}\pm1$ molecules in lower states and $n_{\gamma}\pm1$ photons since they result from the transformation of $\Ket{u_p}$ and $\Ket{u_n}$, respectively, under the action of $\hat{W}$. It follows that the associated molecular states are of the types $\Ket{m\pm1}$ (i.e., $N_g\mp1$ molecules in lower states).

The solution for equation (\ref{eq:c2(t)_again}) is rather lengthy and detailed in Appendix \ref{app:scattering}, but we highlight here a key component resulting from the underlying analysis. That is, the second-order nature of the problem and, accordingly, the repeated appearance of the interaction term $\hat{W}$ in the equation (\ref{eq:c2(t)_again}) brings in double summations on the molecular population of the type
\begin{align}
    \sum_{\alpha,\beta}e^{i\mathbf{k}_{r}\cdot\mathbf{r}_{\beta}}e^{-i\mathbf{k}_{q}\cdot\mathbf{r}_{\alpha}} = \sum_{\alpha}e^{-i\left(\mathbf{k}_{q}-\mathbf{k}_{r}\right)\cdot\mathbf{r}_{\alpha}}+\sum_{\beta\neq\alpha}e^{i\mathbf{k}_{r}\cdot\mathbf{r}_{\beta}}e^{-i\mathbf{k}_{q}\cdot\mathbf{r}_{\alpha}},\label{eq:sum_molecules}
\end{align}
\noindent where $\alpha$ and $\beta$ label the molecules while $q$ and $r$ are for the radiation modes. Upon taking the expectation value of this expression the last term on the right-hand side vanishes because of the random positioning of the molecules. The fact that only the single summation on $\alpha$ remains has for a consequence  that the two transitions in equation (\ref{eq:c2(t)_again}) (through $W_{pj}$ and $W_{kn}$) must happen in only one molecule. Furthermore, it is then possible to show that 
\begin{align}
    \left<e^{-i\left(\mathbf{k}_{q}-\mathbf{k}_{r}\right)\cdot\mathbf{r}_{\alpha}}\right> & = e^{-i\frac{1}{2}\Delta k_{qr}L} \mathrm{sinc}\left(\frac{1}{2}\Delta k_{qr}L\right)\frac{2 J_{1}\left(\frac{1}{2}k_{q}L\sin\theta\right)}{\frac{1}{2}k_{q}L\sin\theta},\label{eq:sum_molecules_2}
\end{align}
\noindent where $\Delta k_{qr}=k_q-k_r$, $\mathrm{sinc}\left(x\right)=\sin\left(x\right)/x$ and $J_{1}\left(x\right)$ is the Bessel function of the first kind of order one. The right-hand side of this equation is strongly peaked at unity and further implies that $\mathbf{k}_{q}=\mathbf{k}_{r}$. As discussed in Appendix \ref{app:scattering}, these considerations force us to conclude that the initial and final states are the same (i.e., $\Ket{u_p}=\Ket{u_n}$) and we are consequently dealing with forward scattering only. 

It follows from this that the state of the system when accounting for resonant scattering (including the zeroth-order term while the first-order term was shown to vanish in Sec. \ref{sec:absorption}) can be written as
\begin{align}
    & \Ket{\psi\left(t\right)} \simeq c_p e^{-iE_{p}^{0}t/\hbar}\left\{1+\frac{it}{\hbar}\frac{3\pi c^{3}}{2\omega_{0}^{3}}A_{eg}u\left(\omega_q\right)\left|\mathbf{\boldsymbol{\epsilon}}_{d}\cdot\boldsymbol{\epsilon}_{q}\right|^{2}N_g\right.\nonumber\\
    & \left.\rule{0mm}{6mm}\times\left[\int\frac{h\left(\omega_\alpha\right)d\omega_\alpha}{\omega_\alpha^+-\omega_q-i\Gamma/2}-\frac{g_g}{g_e}\frac{N_e}{N_g}\int\frac{h\left(\omega_\alpha\right)d\omega_\alpha}{\omega_\alpha^- -\omega_q+i\Gamma/2}\right]\right\}\Ket{u_p},\label{eq:|psi'_p>-2}
\end{align}
\noindent where we, once again, only consider radiation modes at frequency $\omega_q$ directed at the observer, which we combined within the global phase term in the coefficient $c_p$. In equation (\ref{eq:|psi'_p>-2}) $\omega_\alpha^\pm=\omega_\alpha\pm\Phi/\hbar$ and $\omega_q\simeq\omega_q\left(1-\mathbf{v}_\alpha\cdot\mathbf{e}_{q}/c\right)$, while we also have  
\begin{align}
    \Gamma & = \sum_r\left[\Gamma^{m+1}_\mathrm{em}\left(\omega_r\right)+\Gamma^{m+1}_\mathrm{abs}\left(\omega_r\right)+\Gamma^{m-1}_\mathrm{em}\left(\omega_r\right)+\Gamma^{m-1}_\mathrm{abs}\left(\omega_r\right)\right]\nonumber\\
    & \simeq 2\left(1+\frac{g_g}{g_e}\frac{N_e}{N_g}\right)\sum_r\Gamma^m_\mathrm{abs}\left(\omega_r\right)\label{eq:Gamma}\\
    \Phi & = -N_g\left(1-\frac{g_g}{g_e}\frac{N_e}{N_g}\right)\frac{3\pi c^{3}}{\omega_{0}^{3}}A_{eg}\sum_r u\left(\omega_{r}\right)\left|\mathbf{\boldsymbol{\epsilon}}_{d}\cdot\boldsymbol{\epsilon}_{r}\right|^2\nonumber\\
    & \quad\times \int d\omega_\alpha\mathcal{P}\frac{h\left(\omega_\alpha\right)}{\omega_\alpha-\omega_r}.\label{eq:Phi}
\end{align}
\noindent The damping term $\Gamma$ is the sum of all radiation rates out of the virtual states (compare with equations (\ref{eq:Gamma_abs_a})-(\ref{eq:Gamma_se_a})), while $\Phi\left(\ll\hbar\omega_\alpha\right)$ is responsible for a shift in unperturbed energy levels with $\mathcal{P}$ standing for Cauchy's principal value \citep{Cohen-Tannoudji2018b}. This last quantity will have no influence on the results to be presented later on. 

We thus find from equation (\ref{eq:|psi'_p>-2}) that the resonant scattering process is dominated by forward scattering, which is coherent in nature. This is due to the collective behaviour between the large number of molecules acting as a diffraction ensemble that strongly favours this type of scattering over any other mode (see equation (\ref{eq:sum_molecules_2})). This is radically different from the result obtained for absorption/emission process in Sec. \ref{sec:absorption} for which the probability amplitude cancels out due to its non-coherent nature.
 
The picture given by equation (\ref{eq:|psi'_p>-2}) for resonant scattering is thus quite clear: the modes of radiation present in the initial state $\Ket{u_{p}}=\Ket{m}\Ket{n_p}$ pointing in the direction to the observer contribute to the probability amplitude through a term of the type $i\phi_{p}\left(\omega_{q}\right)$ with
\begin{align}
\phi_{p}\left(\omega_{q}\right) = & \frac{t}{\hbar}\frac{3\pi c^{3}}{2\omega_{0}^{3}}A_{eg}u\left(\omega_q\right)\left|\mathbf{\boldsymbol{\epsilon}}_{d}\cdot\boldsymbol{\epsilon}_{q}\right|^{2}N_g\nonumber\\
    & \hspace{-9mm}\rule{0mm}{6mm}\times\left[\int\frac{h\left(\omega_\alpha\right)d\omega_\alpha}{\omega_\alpha^+-\omega_q-i\Gamma/2}-\frac{g_g}{g_e}\frac{N_e}{N_g}\int\frac{h\left(\omega_\alpha\right)d\omega_\alpha}{\omega_\alpha^--\omega_q+i\Gamma/2}\right],\label{eq:phi_p}
\end{align}
\noindent where we should note that $\phi_p$ is a complex quantity in general, especially near resonance when $\omega_q\simeq\omega_\alpha$. 

When $\phi_{p}\left(\omega_{q}\right)\ll 1$ the final state (at that frequency) can be written as
\begin{equation}
\Ket{\psi\left(t\right)}\simeq c_{p}e^{-iE_{p}^{0}t/\hbar}e^{i\phi_{p}\left(\omega_{q}\right)}\Ket{u_{p}},\label{eq:|psi'>_Houde}
\end{equation}
\noindent a form we will adopt later on when modelling polarimetry data.

\subsection{Temporal evolution of probability}

Given equation (\ref{eq:|psi(t)_exp}) the probability of having the system in state $\Ket{u_p}$ at time $t$ is
\begin{equation}
    \mathcal{P}_p\left(t\right) = c_p^2 \left[1+\frac{\left|c^{\left(1\right)}_p\left(t\right)\right|^2}{c_p^2}+\frac{2}{c_p}\mathrm{Re}\left\{c^{\left(2\right)}_p\left(t\right)e^{iE_{p}^{0}t/\hbar}\right\}\right]\label{eq:Prob_p}
\end{equation}
\noindent since the expectation value of $c_p^{\left(1\right)}\left(t\right)$ was shown to vanish in Sec. \ref{sec:absorption} and where $c_p$ was assumed real, while $\mathrm{Re}\left\{\ldots\right\}$ stands for the real part. Using equation (\ref{eq:|psi'_p>-2}) we find
\begin{align}
    \frac{2}{c_p}\mathrm{Re}\left\{c^{\left(2\right)}_p\left(t\right)e^{iE_{p}^{0}t/\hbar}\right\} & \simeq -\frac{t}{\hbar}\frac{3\pi c^{3}}{2\omega_{0}^{3}}A_{eg} u\left(\omega_q\right)\left|\mathbf{\boldsymbol{\epsilon}}_{d}\cdot\boldsymbol{\epsilon}_{q}\right|^{2}N_g\Gamma t\nonumber\\
    & \hspace{-27mm}\times\left[ \int\frac{h\left(\omega_\alpha\right)d\omega_\alpha}{\left(\omega_\alpha^+-\omega_q\right)^2+\Gamma^2/4} + \frac{g_g}{g_e}\frac{N_e}{N_g}\int\frac{h\left(\omega_\alpha\right)d\omega_\alpha}{\left(\omega_\alpha^--\omega_q\right)^2+\Gamma^2/4}\right]\nonumber\\
    & \hspace{-27mm}\simeq -\Gamma_p^{-}\left(\omega_q\right)t\label{eq:RealPart}
\end{align}
\noindent from equations (\ref{eq:Gamma-a})-(\ref{eq:Gamma_se_a}).

On the other hand, we also have (from equations (\ref{eq:|psi'>_abs-1}) and (\ref{eq:Gamma+a}))
\begin{align}
    \frac{\left|c^{\left(1\right)}_p\left(t\right)\right|^2}{c_p^2} & \simeq \left[\Gamma_\mathrm{em}^{m+1}\left(\omega_q\right)+\Gamma_\mathrm{abs}^{m-1}\left(\omega_q\right)\right]t\nonumber\\
    & \simeq \Gamma_p^{+}\left(\omega_q\right)t\label{eq:cp(1)^2}
\end{align}
\noindent such that
\begin{align}
    \mathcal{P}_p\left(t\right) & \simeq c_p^2 \left\{1-\left[\Gamma_{p}^{-}\left(\omega_q\right)-\Gamma_{p}^{+}\left(\omega_q\right)\right]t\right\}\nonumber\\
    & \simeq c_p^2 e^{-\left[\Gamma_{p}^{-}\left(\omega_q\right)-\Gamma_{p}^{+}\left(\omega_q\right)\right]t},\label{eq:Prob_tot}
\end{align}
\noindent where, as could have been expected, the negative of the exponent (divided by the time interval $t$) is the decay rate of state $\Ket{u_p}$ corrected for that due to its gains.

\section{Generation and transformation of polarisation signals through ARS}\label{sec:ARS}


The discussion in the previous section was kept as general as possible in the sense that we did not specify a basis with which to express the polarisation state of the incident electric field in the plane of the sky. There are different sets of bases available to express the polarisation state of the radiation field. For example, the linear basis $\left\{\Ket{n_{p,\Vert}},\Ket{n_{p,\bot}}\right\}$ of polarisation states respectively parallel and perpendicular to the orientation of the projection of the magnetic field on the plane of the sky, or the basis
\begin{align}
    & \Ket{n_{p,R}} = -\frac{1}{\sqrt{2}}\left(\Ket{n_{p,\Vert}} +i\Ket{n_{p,\bot}}\right) \label{eq:n_R}\\
    & \Ket{n_{p,L}} = \frac{1}{\sqrt{2}}\left(\Ket{n_{p,\Vert}} -i\Ket{n_{p,\bot}}\right) \label{eq:n_L} 
\end{align}
\noindent for right and left circular polarisation states, respectively, can be used to decompose a general radiation state $\Ket{n_p}$\footnote{To avoid any confusion we will from now on explicitly write the polarisation state (apart from the mode $p$, here), the understanding being that all radiation modes contained in, say, $\Ket{n_{p,\Vert}}$ are in the ``$\Vert$'' polarisation state, etc.}. Although the linear polarisation basis would intuitively appear to be best suited to situations where the magnetic field is located in the plane of the sky and the circular polarisation basis when it is aligned with the line of sight, it is a priori not clear what basis should be used for an arbitrary magnetic field orientation. However, one can surmise that it will be a function of the coordinate system underlying the analysis. Furthermore, it is also important to realise that the form of our solution, as expressed in equation (\ref{eq:|psi'>_Houde}), not only requires that we somehow choose a basis well suited to our problem but that we find the exact eigen-basis corresponding to the total Hamiltonian (i.e., including the interaction term $\hat{W}$).

\begin{figure}
\begin{center}\includegraphics[scale=0.9]{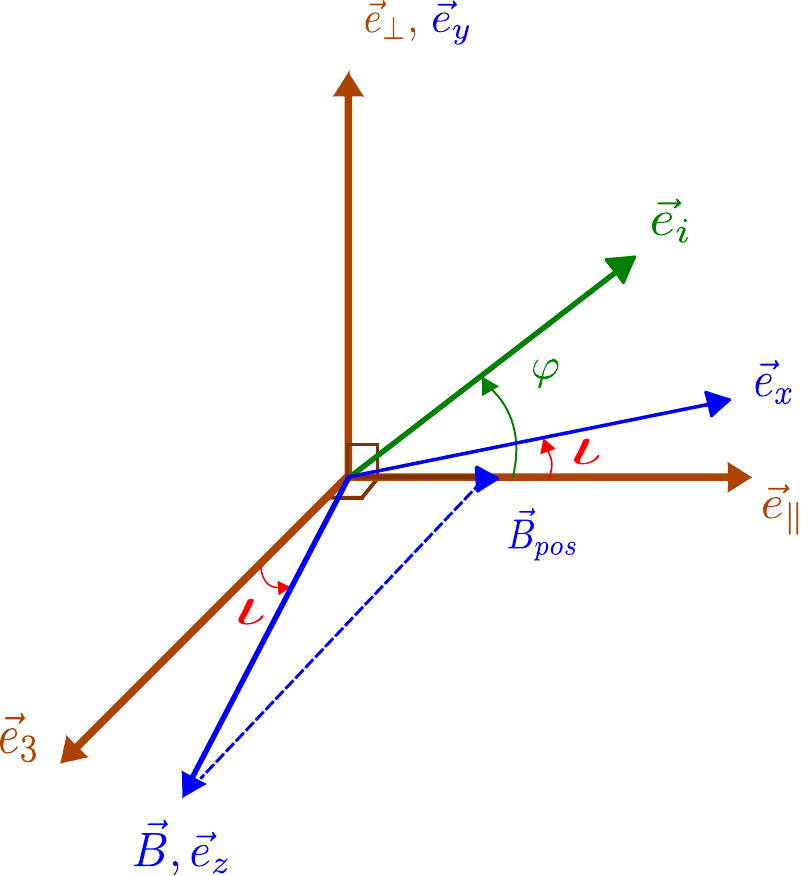}\end{center}
\caption{\label{fig:ARS_coordinates}Coordinate systems. The fixed observer frame of reference is composed of the unit vectors $\mathbf{e}_3$ (along the radiation propagation direction and anti-parallel to the line of sight) and $\left\{\mathbf{e}_\Vert,\mathbf{e}_\bot\right\}$ for the plane of the sky, with $\mathbf{e}_\Vert$ defining the orientation of the projected magnetic field on the plane of the sky. The coordinate system $\left\{\mathbf{e}_x,\mathbf{e}_y,\mathbf{e}_z\right\}$ defines the molecular orientation, with the magnetic field pointing along $\mathbf{e}_z$ at an inclination angle $\iota$ from $\mathbf{e}_3$; the unit vectors $\mathbf{e}_3,\,\mathbf{e}_z,\,\mathbf{e}_\Vert$, and $\mathbf{e}_x$ all share the same plane. The incident electric field is polarised along $\mathbf{e}_i$, at an angle $\varphi$ from $\mathbf{e}_\Vert$.}
\end{figure}

Our choice of coordinate systems is presented in Figure \ref{fig:ARS_coordinates}. The system adopted for the fixed observer frame is composed of the unit vectors $\mathbf{e}_3$ for the direction of radiation propagation (and anti-parallel to the line of sight) and $\left\{\mathbf{e}_\Vert,\mathbf{e}_\bot\right\}$ for the plane of the sky, with $\mathbf{e}_\Vert$ defining the orientation of the projected magnetic field on the plane of the sky. The other coordinate system $\left\{\mathbf{e}_x,\mathbf{e}_y,\mathbf{e}_z\right\}$ is for the frame defining the molecular orientation, with the magnetic field pointing along $\mathbf{e}_z$ at an inclination angle $\iota$ from $\mathbf{e}_3$. We note that the unit vectors $\mathbf{e}_3,\,\mathbf{e}_z,\,\mathbf{e}_\Vert$, and $\mathbf{e}_x$ all share the same plane, while it is understood that unit vectors and corresponding radiation states share the same subscript, e.g., $\Ket{n_{p,\Vert}}$ is for a polarisation state along $\mathbf{e}_\Vert$, etc.

If we focus our analysis on a simple linear molecule such as CO, then the basis made of the polarisation states $\Ket{n_{p,z}}$ and
\begin{align}
    & \Ket{n_{p,+}} = -\frac{1}{\sqrt{2}}\left(\Ket{n_{p,x}}+i\Ket{n_{p,y}}\right) \label{eq:n_+} \\
    & \Ket{n_{p,-}} = \frac{1}{\sqrt{2}}\left(\Ket{n_{p,x}}-i\Ket{n_{p,y}}\right) \label{eq:n_-}
\end{align}
follows naturally for describing the radiation stemming from the $\pi$, $\sigma_+$ and $\sigma_-$ transitions, respectively (note that $\Ket{n_{p,y}}=\Ket{n_{p,\bot}}$). Considering the symmetries imposed on the system by the presence of the magnetic field, it is shown in Appendix \ref{app:elliptical} that the appropriate two-dimensional eigen-basis to describe the radiation polarisation state in the observer's frame for our analysis is
\begin{align}
    & \Ket{n_{p,1}} = \frac{-1}{\sqrt{1+\cos^2\iota}}\left(\cos\iota\Ket{n_{p,\Vert}}+i\Ket{n_{p,\bot}}\right)\label{eq:n_1}\\
    & \Ket{n_{p,2}} = \frac{1}{\sqrt{1+\cos^2\iota}}\left(\Ket{n_{p,\Vert}}-i\cos\iota\Ket{n_{p,\bot}}\right)\label{eq:n_2}.
\end{align}
\noindent Equations (\ref{eq:n_1})-(\ref{eq:n_2}) define an elliptical polarisation basis whose eccentricity is set by $\iota$, the angle of inclination of the magnetic field to (minus) the line of sight. We can verify that $\left\{\Ket{n_{p,1}},\Ket{n_{p,2}}\right\}$ tends to the linear polarisation basis when $\iota\rightarrow\pi/2$, and to the circular polarisation basis when $\iota\rightarrow 0$ or $\pi$ (in these cases the orientation of $\mathbf{e}_\Vert$ becomes arbitrary). 

Although the analysis to follow can be readily applied to any incoming radiation polarisation state, we will focus on the case of an incident linear polarisation signal at an angle $\varphi$ from the $\mathbf{e}_\Vert$-axis, as shown in Figure \ref{fig:ARS_coordinates}. That is, the corresponding incident polarisation state is
\begin{equation}
    \Ket{n_{p,i}} = \cos\varphi\Ket{n_{p,\Vert}}+\sin\varphi\Ket{n_{p,\bot}},\label{eq:|n_i>}
\end{equation}
\noindent which we expand using equations (\ref{eq:n_1})-(\ref{eq:n_2}) to
\begin{align}
    \Ket{n_{p,i}} & = \frac{1}{\sqrt{1+\cos^2\iota}}\left[\left(-\cos{\iota}\cos{\varphi}+i\sin{\varphi}\right)\Ket{n_{p,1}}\right.\nonumber\\
    & \qquad\qquad\qquad\left.+\left(\cos{\varphi}+i\cos\iota\sin\varphi\right)\Ket{n_{p,2}}\right].\label{eq:|n_i>_2}
\end{align}

With the corresponding elliptical unit vectors $\left\{\mathbf{e}_1,\mathbf{e}_2\right\}$ and $\left\{\mathbf{e}_z,\mathbf{e}_+,\mathbf{e}_-\right\}$ for the $\pi$, $\sigma_+$ and $\sigma_-$ lines, as well as the equation defining the electric dipole moment (see Appendix \ref{app:elliptical}) 
\begin{align}
    \hat{\mathbf{d}} = \hat{d}_\pi\mathbf{e}_{\pi}+\hat{d}_+\mathbf{e}_{+}^*+\hat{d}_-\mathbf{e}_{-}^*,\label{eq:d_main_text}
\end{align}
we can calculate the relative phase shifts accrued through resonant scattering between the two elliptical polarisation states associated with these transitions (using equation (\ref{eq:phi_p})). That is, if we first define for the radiation state $\Ket{n_p}$ (i.e., for modes pointing at the observer) at frequency $\omega$ (we drop the frequency subscript $q$ from now on) 
\begin{align}
    \Delta\phi_{p,\pi} & \left(\omega\right) = \frac{t}{\hbar}\frac{3\pi c^{3}}{2\omega_0^3}A_{eg}u\left(\omega\right)\left(\left|\mathbf{e}_z\cdot\mathbf{e}_1\right|^2-\left|\mathbf{e}_z\cdot\mathbf{e}_2\right|^{2}\right)I_\pi\left(\omega\right)\label{eq:phi_p_pi}\\
    \Delta\phi_{p,\pm} & \left(\omega\right) = \frac{t}{\hbar}\frac{3\pi c^{3}}{2\omega_0^3}A_{eg}u\left(\omega\right)\left(\left|\mathbf{e}_\pm^*\cdot\mathbf{e}_1\right|^2-\left|\mathbf{e}_\pm^*\cdot\mathbf{e}_2\right|^{2}\right)I_\pm\left(\omega\right)\label{eq:phi_p_pm}
\end{align}
\noindent and 
\begin{align}
    I_\pi\left(\omega\right) & = N_g\int\frac{h\left(\omega_\alpha\right)d\omega_\alpha}{\omega_\alpha^+-\omega-i\Gamma/2}-\frac{g_g}{g_e}N_e\int\frac{h\left(\omega_\alpha\right)d\omega_\alpha}{\omega_\alpha^--\omega+i\Gamma/2}\label{eq:I_pi}\\
    I_\pm\left(\omega\right) & = N_g\int\frac{h\left(\omega_\alpha\right)d\omega_\alpha}{\omega_\alpha^+\pm\omega_z-\omega-i\Gamma/2}\nonumber\\
    & -\frac{g_g}{g_e}N_e\int\frac{h\left(\omega_\alpha\right)d\omega_\alpha}{\omega_\alpha^-\pm\omega_z-\omega+i\Gamma/2}\label{eq:I_+-}
\end{align}
\noindent with $\omega_z$ the Zeeman splitting, then the total relative phase shift is given by
\begin{align}
    \Delta\phi_{p}\left(\omega\right) & = \Delta\phi_{p,\pi}\left(\omega\right)+\Delta\phi_{p,+}\left(\omega\right)+\Delta\phi_{p,-}\left(\omega\right)\nonumber\\
    & = \Delta\phi_{p,\mathrm{s}}\left(\omega\right)+\Delta\phi_{p,\mathrm{c}}\left(\omega\right)\label{eq:Dphi_p}
\end{align}
\noindent where 
\begin{align}
    \Delta \phi_{p,\mathrm{s}}\left(\omega\right) & = -\frac{t}{\hbar}\frac{3\pi c^{3}}{2\omega_0^3}A_{eg}u\left(\omega\right)\nonumber\\
    & \times\frac{\sin^4\iota}{1+\cos^2\iota}\left\{I_\pi\left(\omega\right)-\frac{1}{2}\left[I_{+}\left(\omega\right)+I_{-}\left(\omega\right)\right]\right\}\label{eq:Dphi_p,s}\\
    \Delta \phi_{p,\mathrm{c}}\left(\omega\right) & = \frac{t}{\hbar}\frac{3\pi c^{3}}{2\omega_0^3}A_{eg}u\left(\omega\right)\frac{2\cos^2\iota}{1+\cos^2\iota}\left[I_{+}\left(\omega\right)-I_{-}\left(\omega\right)\right].\label{eq:Dphi_p,c}
\end{align}


Whenever $\omega_z$ is much smaller than $\Gamma$ and the width of $h\left(\omega_\alpha\right)$ it can be shown that (by expanding the reciprocal of the denominators in the integrands of equation (\ref{eq:I_+-}) with second-order Taylor series about $\omega_\alpha^\pm$ and integrating by parts) 
\scriptsize
\begin{align}
    I_\pi\left(\omega\right) & -\frac{1}{2}\left[I_{+}\left(\omega\right)+I_{-}\left(\omega\right)\right] \simeq -\frac{\omega_z^2}{2}\nonumber\\
    & \times\left[ N_g\int\frac{d^2 h\left(\omega_\alpha\right)/d\omega_\alpha^2}{\omega_\alpha^+-\omega-i\Gamma/2}d\omega_\alpha-\frac{g_g}{g_e}N_e\int\frac{d^2 h\left(\omega_\alpha\right)/d\omega_\alpha^2}{\omega_\alpha^--\omega+i\Gamma/2}d\omega_\alpha\right]\label{eq:Is(w)_app}\\
    I_{+}\left(\omega\right) & -I_{-}\left(\omega\right) \simeq -2\omega_z\nonumber\\
     & \times\left[ N_g\int\frac{dh\left(\omega_\alpha\right)/d\omega_\alpha}{\omega_\alpha^+-\omega-i\Gamma/2}d\omega_\alpha-\frac{g_g}{g_e}N_e\int\frac{dh\left(\omega_\alpha\right)/d\omega_\alpha}{\omega_\alpha^--\omega+i\Gamma/2}d\omega_\alpha\right].\label{eq:Ic(w)_app}
\end{align}
\normalsize
We therefore find from equations (\ref{eq:Dphi_p})-(\ref{eq:Ic(w)_app}) that the relative phase shift $\Delta\phi_{p}\left(\omega\right)$ is made of contributions from one part necessitating the existence of magnetic field component in the plane of the sky (i.e., $\Delta\phi_{p,\mathrm{s}}$, which is proportional to $\left(\omega_z\sin\iota\right)^2$) and from another requiring a component along the line of sight (i.e., $\Delta\phi_{p,\mathrm{c}}$, which is proportional to $\omega_z\cos\iota$). Although our discussion will be kept as general as possible, it is found that $\Delta\phi_{p,\mathrm{c}}\gg \Delta\phi_{p,\mathrm{s}}$ for physical conditions encountered in most astronomical media relevant to this type of analysis.  We also note that, although it is a complex quantity, the relative phase shift $\Delta\phi_p$ tends to become real far away from resonance (i.e., when $\omega_\alpha^\pm-\omega\gg\Gamma$).

The final, scattered, radiation state at frequency $\omega$ is then given by (see equation (\ref{eq:|psi'>_Houde}))
\begin{align}
    \Ket{n_{p,f}} & = \frac{1}{\sqrt{1+\cos^2\iota}}\left[e^{i\Delta\phi_p\left(\omega\right)}\left(-\cos{\iota}\cos{\varphi}+i\sin{\varphi}\right)\Ket{n_{p,1}}\right.\nonumber\\
    & \qquad\qquad\qquad\left.+\left(\cos{\varphi}+i\cos\iota\sin\varphi\right)\Ket{n_{p,2}}\right].\label{eq:|n_f>}
\end{align}
Using the linear bases $\left\{\Ket{n_{p,\Vert}},\Ket{n_{p,\bot}}\right\}$ and 
\begin{equation}
    \Ket{n_{p,\pm45}} = \frac{1}{\sqrt{2}}\left(\Ket{n_{p,\Vert}}\pm\Ket{n_{p,\bot}}\right),\label{eq:|n_45>}
\end{equation}
\noindent as well as the circular polarisation basis $\left\{\Ket{n_{p,R}},\Ket{n_{p,L}}\right\}$ (see equations (\ref{eq:n_R})-(\ref{eq:n_L})) the normalized Stokes parameters at frequency $\omega$ can be expressed as
\begin{align}
    q_p & = \frac{\left|\Braket{n_{p,\Vert}|n_{p,f}}\right|^2-\left|\Braket{n_{p,\bot}|n_{p,f}}\right|^2}{\left|\Braket{n_{p,1}|n_{p,f}}\right|^2+\left|\Braket{n_{p,2}|n_{p,f}}\right|^2}\label{eq:Stokes_q}\\
    u_p & = \frac{\left|\Braket{n_{p,+45}|n_{p,f}}\right|^2-\left|\Braket{n_{p,-45}|n_{p,f}}\right|^2}{\left|\Braket{n_{p,1}|n_{p,f}}\right|^2+\left|\Braket{n_{p,2}|n_{p,f}}\right|^2}\label{eq:Stokes_u}\\
    v_p & = \frac{\left|\Braket{n_{p,R}|n_{p,f}}\right|^2-\left|\Braket{n_{p,L}|n_{p,f}}\right|^2}{\left|\Braket{n_{p,1}|n_{p,f}}\right|^2+\left|\Braket{n_{p,2}|n_{p,f}}\right|^2}.\label{eq:Stokes_v}
\end{align}
\noindent The normalisation factor 
\begin{equation}
    i_p = \left|\Braket{n_{p,1}|n_{p,f}}\right|^2+\left|\Braket{n_{p,2}|n_{p,f}}\right|^2\label{eq:Stokes_i}
\end{equation}
\noindent is needed in the denominator of equations (\ref{eq:Stokes_q})-(\ref{eq:Stokes_v}) to account for the potential probability decay or enhancement of state $\Ket{n_{p,f}}$ as the system evolves. These equations are expressed as functions of the different parameters in Appendix \ref{app:stokes}. 

To investigate the behaviour of the polarisation signals under different situations it will be advantageous to first express the relative phase shift $\Delta\phi_p$ as a function of its real and imaginary parts with 
\begin{equation}
    \Delta\phi_p = \Delta\eta_p+i\Delta\gamma_p,\label{eq:Delta_phi}
\end{equation}
\noindent where a dependency on the radiation frequency $\omega$ is implied. 

In particular, it can be verified from equations (\ref{eq:Stokes_q_main})-(\ref{eq:Stokes_i_main}) that in the absence of a relative phase shift (i.e., when $\Delta\phi_p=0$) we recover the values 
\begin{align}
    q_p & = \cos(2\varphi)\label{eq:q_0}\\
    u_p & = \sin(2\varphi)\label{eq:u_0}\\
    v_p & = 0\label{eq:v_0}
\end{align}
\noindent expected for the incident linearly polarised state given in equation (\ref{eq:|n_i>}). It is easily verified that this condition corresponds to the case $\omega_z=0$, i.e., when no magnetic field is present. 

Although the expressions for the Stokes parameters are in general quite complex, they greatly simplify when the magnetic field is either parallel or perpendicular to the plane of the sky. We now turn our attention to these special cases. 

\subsection{Conversion between linear and circular polarisation \texorpdfstring{($\iota=\pi/2$)}{}}\label{sec:conversion} 

In the limit where the magnetic field is confined to the plane of the sky, i.e., when $\iota=\pi/2$, the equations derived above greatly simplify. More precisely, taking the limit $\iota\rightarrow\pi/2$ in equations (\ref{eq:Dphi_p})-(\ref{eq:Dphi_p,c}) and (\ref{eq:Stokes_q})-(\ref{eq:Stokes_v}) we have
\begin{equation}
    \Delta\phi_p\left(\omega\right) = -\frac{t}{\hbar}\frac{3\pi c^{3}}{2\omega_0^3}A_{eg}u\left(\omega\right)\left\{I_\pi\left(\omega\right)-\frac{1}{2}\left[I_{+}\left(\omega\right)+I_{-}\left(\omega\right)\right]\right\},\label{eq:Dphi_p_rot}
\end{equation}
\noindent which is proportional to $\omega_z^2$, while from equations (\ref{eq:Stokes_q_C})-(\ref{eq:Stokes_i_C})
\begin{align}
    i_p & = e^{-2\Delta\gamma_p}\sin^2\varphi+\cos^2\varphi\label{eq:i_conv}\\
    q_p & = \left(\cos^2\varphi-e^{-2\Delta\gamma_p}\sin^2\varphi\right)/i_p\label{eq:q_conv}\\
    u_p & = e^{-\Delta\gamma_p}\sin(2\varphi)\cos\left(\Delta\eta_p\right)/i_p\label{eq:u_conv}\\
    v_p & = e^{-\Delta\gamma_p}\sin(2\varphi)\sin\left(\Delta\eta_p\right)/i_p.\label{eq:v_conv}
\end{align}
\noindent In the limit when the loss term is negligible (i.e., $\Delta\gamma_p\rightarrow0$) we have 
\begin{align}
    i_p & = 1\label{eq:i_conv_1}\\
    q_p & = \cos(2\varphi)\label{eq:q_conv_1}\\
    u_p & = \sin(2\varphi)\cos\left(\Delta\eta_p\right)\label{eq:u_conv_1}\\
    v_p & = \sin(2\varphi)\sin\left(\Delta\eta_p\right).\label{eq:v_conv_1}
\end{align}
\noindent This form for the Stokes parameters corresponds to a (Faraday) conversion between linear and circular polarisations (i.e., between Stokes~$u_p$ and Stokes~$v_p$) as a relative phase shift $\Delta\phi_p$ builds up during the evolution of the system. This is the case that was originally considered by \citet{Houde2013}.

\subsection{Faraday rotation \texorpdfstring{($\iota=0$ or $\pi$)}{}}\label{sec:rotation} 

In the opposite limit where the magnetic field is (anti-)aligned with the line of sight, i.e., when $\iota=0$ or $\pi$, 
\begin{equation}
    \Delta\phi_p\left(\omega\right) = \frac{t}{\hbar}\frac{3\pi c^{3}}{2\omega_0^3}A_{eg}u\left(\omega\right)\left[I_{+}\left(\omega\right)-I_{-}\left(\omega\right)\right]\label{eq:Dphi_p_conv}
\end{equation}
\noindent is now proportional to $\omega_z$, and
\begin{align}
    i_p & = e^{-\Delta\gamma_p}\cosh\left(\Delta\gamma_p\right)\label{eq:i_rot}\\
    q_p & = e^{-\Delta\gamma_p}\cos\left(2\varphi\mp\Delta\eta_p\right)/i_p\label{eq:q_rot}\\
    u_p & = e^{-\Delta\gamma_p}\sin\left(2\varphi\mp\Delta\eta_p\right)/i_p\label{eq:u_rot}\\
    v_p & = \mp e^{-\Delta\gamma_p}\sinh\left(\Delta\gamma_p\right)/i_p,\label{eq:v_rot}
\end{align}
where the `-' (`+') sign is for $\iota=0$ ($\pi$). In the limit when $\Delta\gamma_p\rightarrow0$ these equations transform to 
\begin{align}
    i_p & = 1\label{eq:i_rot_1}\\
    q_p & = \cos\left(2\varphi\mp\Delta\eta_p\right)\label{eq:q_rot_1}\\
    u_p & = \sin\left(2\varphi\mp\Delta\eta_p\right)\label{eq:u_rot_1}\\
    v_p & = 0.\label{eq:v_rot_1}
\end{align}
This corresponds to a Faraday rotation of the initial linear polarisation states (i.e., involving Stokes~$q_p$ and Stokes~$u_p$) as a relative phase shift $\Delta\phi_p$ builds up as the system evolves. 

\subsection{Arbitrary magnetic field orientation}\label{sec:arbitrary}

In general, the magnetic field will have an arbitrary inclination angle $\iota$ while the incident electric field can be decomposed into a sum of polarised and unpolarised components. That is, the incident radiation can be written as
\begin{equation}
    \Ket{n_{p,i}} = \left(\alpha_0 + \alpha_1\right)\Ket{n_{p,\Vert}}+\left(\beta_0 + \beta_1\right)\Ket{n_{p,\bot}},\label{eq:pol_unpol}
\end{equation}
\noindent where $\alpha_0$ and $\beta_0$ are for the (deterministic) polarised component, while $\alpha_1$ and $\beta_1$ stand for the (random) unpolarised part. Given its random nature, the unpolarised component of the state has the following property for its mean values $\left<\alpha_1\right>=\left<\beta_1\right>=\left<\alpha_1\beta_1\right>=0$. It therefore follows that only terms in the form of $\left<\alpha_1^2\right>$ and $\left<\beta_1^2\right>$ can contribute to the Stokes parameters. This does not apply to the polarised part of the state. 

Given these properties, the polarised and unpolarised parts of the radiation state can be treated independently, with their contributions to the Stokes parameters combined afterwards to determine the final polarisation state of the emerging radiation (see Appendix \ref{app:stokes}).

\subsubsection{Generation of polarisation signals}\label{sec:generation}

To deal with an upolarised incident radiation state we set $\alpha_0=\beta_0=0$, $\alpha_1=\cos\left(\varphi_1\right)$  and $\beta_1=\sin\left(\varphi_1\right)$ with the angle $\varphi_1$ a random variable such that, as mentioned before, $\left<\alpha_1\right>=\left<\beta_1\right>=\left<\alpha_1\beta_1\right>=0$ while $\left<\alpha_1^2\right>=\left<\beta_1^2\right>=1/2$. It is then found from equations (\ref{eq:Stokes_i_C})-(\ref{eq:Stokes_v_C}) that the normalised Stokes parameters pertaining to the emerging radiation state simplify to
\begin{align}
    \left<q_p\right> & = \frac{\sin^2\iota}{1+\cos^2\iota}\tanh{\left(\Delta\gamma_p\right)}\label{eq:qp_unp}\\
    \left<u_p\right> & = 0\label{eq:up_unp}\\
    \left<v_p\right> & = \frac{-2\cos\iota}{1+\cos^2\iota}\tanh{\left(\Delta\gamma_p\right)},\label{eq:vp_unp}
\end{align}
\noindent where a dependency on the radiation frequency $\omega$ is again implied (through $\Delta\gamma_p$). 

We therefore find from equations (\ref{eq:qp_unp})-(\ref{eq:vp_unp}) the significant result that the ARS effect will produce elliptical polarisation signals from an incident unpolarised radiation field. More precisely, we find a linear polarisation signature aligned with or perpendicular to the plane of the sky component of the magnetic field (sharing this ambiguity with the Goldreich-Kylafis effect; see \citealt{Goldreich1981}) as well as a circular polarisation part. The relative strength of the linear and circular polarisation signals is determined by the inclination angle $\iota$ of the magnetic field while the total amount of polarisation is set by the parameter $\Delta\gamma_p$. It is, however, important to note that $\Delta\gamma_p$ also has a dependency on $\iota$ through equations (\ref{eq:Dphi_p})-(\ref{eq:Dphi_p,s}) and (\ref{eq:Delta_phi}). 

Under the assumption that all polarisation signals are generated in the manner discussed here (i.e., there is no transformation of polarised incident radiation, as in Sec. \ref{sec:transformation} below, for example) the main parameters characterizing the polarisation states in equations (\ref{eq:qp_unp})-(\ref{eq:vp_unp}) can, in principle, be directly determined from observations, albeit with some constraints. More precisely, we find that 
\begin{align}
    & \sin{\iota} = \left[2\frac{\left<q_p\right>^2}{\left<v_p\right>^2}\left(\sqrt{1+\frac{\left<v_p\right>^2}{\left<q_p\right>^2}}-1\right)\right]^{1/2}\label{eq:tan_iota}\\
    & \tanh{\left(\Delta\gamma_p\right)} = \pm\sqrt{\left<q_p\right>^2+\left<u_p\right>^2+\left<v_p\right>^2.}\label{eq:tanh}
\end{align}
\noindent We therefore also find an ambiguity in the inclination angle $\iota$ from equation (\ref{eq:tan_iota}) since $\sin\iota = \sin\left(\pi-\iota\right)$, not unlike what is seen for the orientation of the linear polarisation state on the plane of the sky. Combined with the sign uncertainty present in equation (\ref{eq:tanh}), we are left with four different available combinations to fit observational data. 

Finally, we note that while the sign of $q_p$ and equation (\ref{eq:tan_iota}) informs us on the orientation of the magnetic field in space, equation (\ref{eq:tanh}) does the same for its strength through equations (\ref{eq:Dphi_p})-(\ref{eq:Ic(w)_app}) and (\ref{eq:Delta_phi}). 

\subsubsection{Transformation of polarisation}\label{sec:transformation}

In order to get a sense of the transformation of an incident linear polarisation signal, as defined in equation (\ref{eq:|n_i>}), we proceed by approximating $\Delta\phi_p\simeq\Delta\eta_p$. In other words, we neglect the imaginary part of the relative phase shift that is responsible for the generation of polarisation signals dealt with in the previous section (i.e., we set $\Delta\gamma_p=0$ in equation (\ref{eq:Delta_phi})). Under such circumstances the Stokes parameters simplify to (see equations (\ref{eq:Stokes_q_main})-(\ref{eq:Stokes_i_main}))
\begin{align}
    q_p & \simeq \frac{1}{\left(1+\cos^2\iota\right)^2}\cos\left(2\varphi\right)\left[\sin^4\iota+4\cos^2\iota\cos\left(\Delta\eta_p\right)\right]\nonumber\\
    & \quad +\frac{2\cos\iota}{1+\cos^2\iota}\sin\left(2\varphi\right)\sin\left(\Delta\eta_p\right)\label{eq:Stokes_q_pol}\\
    u_p & \simeq \sin\left(2\varphi\right)\cos\left(\Delta\eta_p\right)-\frac{2\cos\iota}{1+\cos^2\iota}\cos\left(2\varphi\right)\sin\left(\Delta\eta_p\right)\label{eq:Stokes_u_pol}\\
    v_p & \simeq \frac{\sin^2\iota}{1+\cos^2\iota}\left\{\rule{0mm}{5mm}\sin\left(2\varphi\right)\sin\left(\Delta\eta_p\right)\right.\nonumber\\
    & \quad\left.+\frac{2\cos\iota}{1+\cos^2\iota}\cos\left(2\varphi\right)\left[\cos\left(\Delta\eta_p\right)-1\right]\right\}.\label{eq:Stokes_v_pol}
\end{align}

Given that the polarisation state of the incident radiation is characterised equations by (\ref{eq:q_0})-(\ref{eq:v_0}), we see that the inclination angle brings an interaction between the different Stokes parameters. In general, the first terms on the right-hand side of these equations are responsible for the Faraday conversion effect discussed in Sec. \ref{sec:conversion} while the last terms bring in the Faraday rotation of Sec. \ref{sec:rotation}. At intermediate values for the inclination angle $\iota$ we have a mixture of the two effects, as a phase shift builds up with the temporal evolution of the system.   

\section{Application to astronomical data}\label{sec:application}

We now turn our attention to the application of our ARS model to astronomical data. To do so we must first express key parameters in term of observable or measurable quantities. For example, the total number of molecules in the lower states $N_g$ depends on the molecular density, the level of excitation (i.e., the temperature $T$ under thermodynamic equilibrium) and the volume of quantization $L^3$. In turn, the volume's linear size is related to the relaxation rate $\Gamma$ though
\begin{align}
    L = \frac{c}{\Gamma}\label{eq:L}
\end{align}
\noindent such that
\begin{align}
    N_g & = n_g L^3\nonumber\\
    & = n\frac{g_g e^{-E_g/kT}}{Q\left(T\right)}\left(\frac{c}{\Gamma}\right)^3,\label{eq:Ng}
\end{align}
\noindent with $n$ the molecular density and $Q\left(T\right)$ the partition function at temperature $T$. Also, the energy density of radiation at frequency $\omega$ is given by
\begin{align}
    u\left(\omega\right) = F\left(\omega\right)\frac{\Gamma}{c},\label{eq:u(w)}
\end{align}
\noindent where $F\left(\omega\right)$ is the corresponding measured flux density.

Since we also know from equation (\ref{eq:Gamma}) that $\Gamma\simeq 4\,\Gamma_\mathrm{abs}^m\left(\omega\right)$ we can write, using equations (\ref{eq:Gamma_abs_a}) and (\ref{eq:L})-(\ref{eq:u(w)}),
\begin{align}
    \Gamma^3 \simeq n\frac{g_g e^{-E_g/kT}}{Q\left(T\right)}\frac{12\pi^{2}c^{5}}{\hbar\omega_{0}^{3}}A_{eg}F\left(\omega\right)h\left(\omega\right).\label{eq:Gamma^3}
\end{align}

\begin{figure}
\begin{center}\includegraphics[scale=0.45]{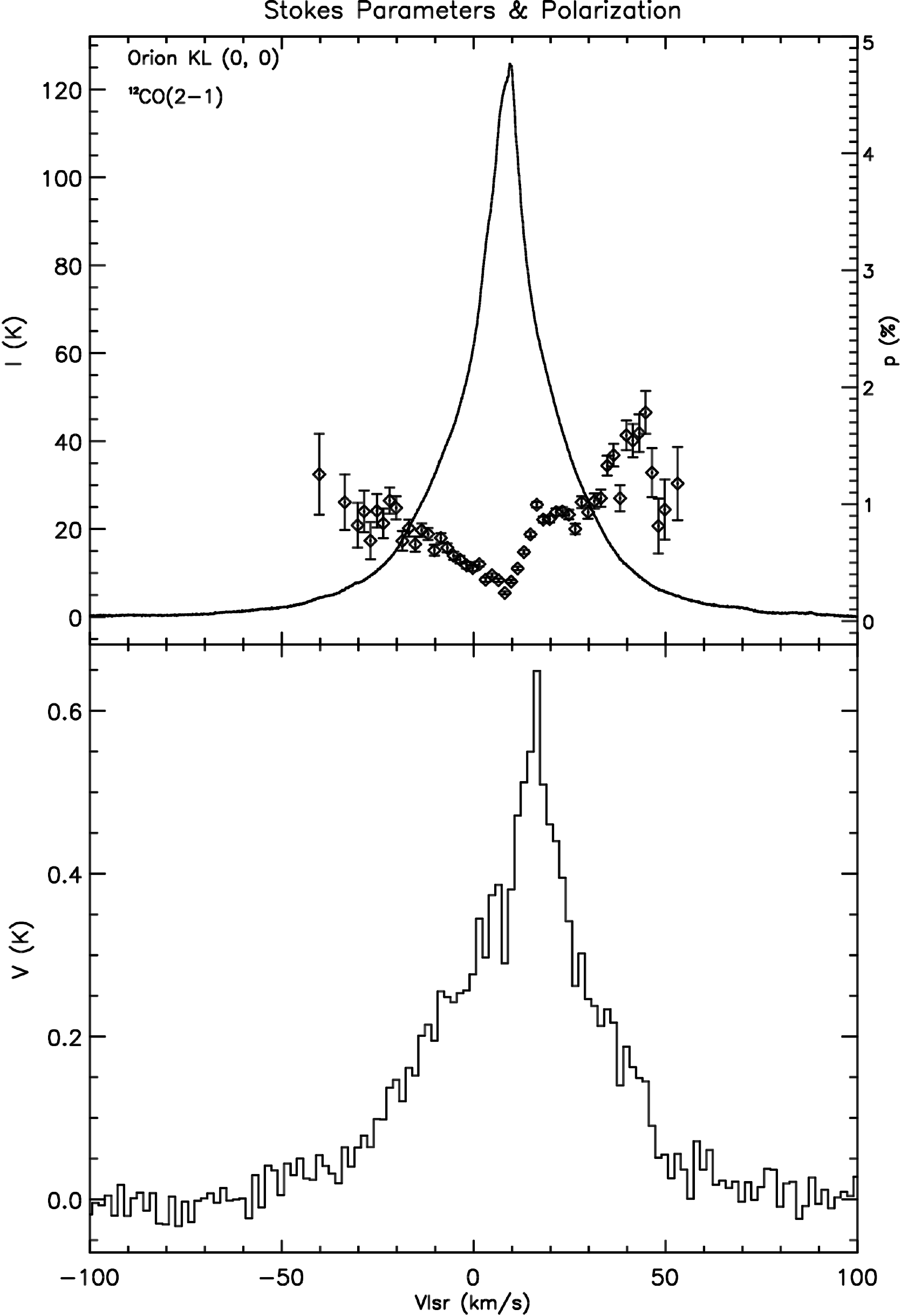}\end{center}
\caption{\label{fig:OrionKL_cp}Discovery spectrum of circular polarisation in $\mathrm{CO}\left(J=2\rightarrow 1\right)$ at the peak intensity position of Orion KL. \textit{Top panel:} Stokes $I$ spectrum, uncorrected for telescope efficiency, with the percentage of polarisaton (symbols; relative to Stokes $I$) using the vertical scale on the right. \textit{Bottom panel:} Stokes $V$ spectrum. Taken from \citet{Houde2013}.}
\end{figure}

\begin{figure}
\begin{center}\includegraphics[scale=0.45]{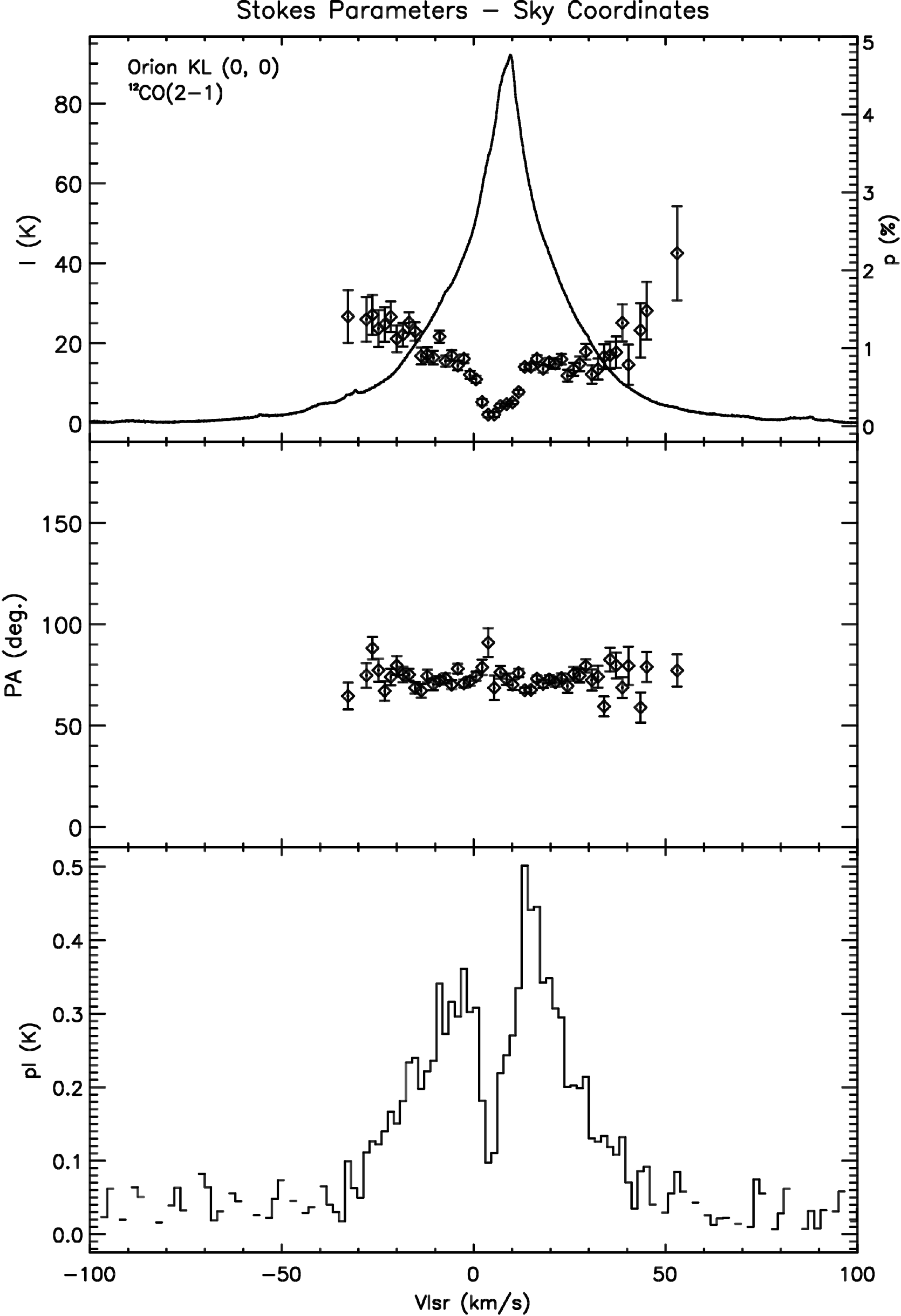}\end{center}
\caption{\label{fig:OrionKL_lp}Linear polarisation spectrum in $\mathrm{CO}\left(J=2\rightarrow 1\right)$ at the peak intensity position of Orion KL. \textit{Top panel:} Stokes $I$ spectrum, uncorrected for telescope efficiency, with the percentage of polarisaton (symbols; relative to Stokes $I$) using the vertical scale on the right. Note that the difference in peak intensity compared to Figure \ref{fig:OrionKL_cp} is due to the fact that the corresponding observations were conducted on different days, with a slight offset in pointing being its cause \citep{Houde2013}. \textit{Middle panel:} Polarisation angle (PA) defined relative to north and increasing eastward. \textit{Bottom panel:} Linear polarisation intensity spectrum. Taken from \citet{Houde2013}.}
\end{figure}

We now revisit the discovery spectrum of circular polarisation in $\mathrm{CO}\left(J=2\rightarrow 1\right)$ at the peak intensity position of Orion KL obtained with the Four-Stokes-Parameter spectral line Polarimeter (FSPPol; \citealt{Hezareh2010}) at the Caltech Submillimeter Observatory (CSO) and originally published by \citet{Houde2013}. The corresponding circular and linear polarisation spectra are shown in Figures \ref{fig:OrionKL_cp} and \ref{fig:OrionKL_lp}. As discussed in \citet{Houde2013}, instrumental polarisation (IP) levels on the order of $0.2-0.3\%$ are expected for FSPPol \citep{Hezareh2010}, which are consistent with the levels detected at the position of peak intensity in Figures \ref{fig:OrionKL_cp} and \ref{fig:OrionKL_lp}. We note that the corresponding polarisation spectra have not been corrected for IP. This should be kept in mind since contamination from telescope side-lobes is likely to play a role for polarisation measurements on extended sources, such as OMC-1. It is therefore likely that there exists some contribution from IP in the observations presented here.

Following \citet{Houde2013} we use the model of \citet{Plume2012} who estimated from CO isotopologues a number density $n_\mathrm{H_2}=10^7\,\mathrm{cm}^{-3}$ and an excitation temperature $T=150$~K for the Orion KL Hot Core. Assuming a CO abundance of $\chi_\mathrm{CO}=10^{-4}$, we have $n=10^3\,\mathrm{cm}^{-3}$, $Q\left(T\right)=54.6$ and $g_g n_e/\left(g_e n_g\right)\simeq 0.93$ ($g_g=3$ and $g_e=5$). The flux density $F\left(\omega\right)$ comprises two components: one for the Stokes $I$ CO spectral line (peaking at $T_A^*\simeq 60$~K for one polarisation from Figure \ref{fig:OrionKL_cp}) and another from the continuum radiation at the frequency of the $\mathrm{CO}\left(J=2\rightarrow 1\right)$ transition (i.e., 230.5~GHz). Using a peak flux of $\simeq 20$~Jy at 850~$\micron$ in a beam of $14\arcsec$ from \citet{Johnstone1999} as a starting point, we estimate a continuum contribution of approximately $1.6$~K in the $32\arcsec$ CSO beam at that frequency (one polarisation). The total flux density can then be converted to Jy (the corresponding factor is $32.5\;\mathrm{Jy/K}$ for the CSO, while the telescope efficiency was $\sim 0.6$). Other relevant parameters are $A_{eg}= 7\times 10^{-7}\,\mathrm{s}^{-1}$, $E_g/k=5.5$~K, $E_e/k=16.6$~K and a Land\'{e} factor $g_J^\mathrm{CO}=-0.296$ ($\omega_z=g_J^\mathrm{CO}B$; \citealt{Gordy1984}). Finally, as will be discussed below we model the scattering molecular population using a Gaussian velocity distribution with a peak value $h\left(\omega_0\right) = 1/\left(\sqrt{2\pi}\Delta\omega\right)\simeq 4\times10^{-9}\,\mathrm{rad\,s}^{-1}$ (i.e., a velocity distribution of $c\Delta\omega/\omega_0 \simeq 20\,\mathrm{km\,s}^{-1}$, corresponding to  a FWHM of approximately $46\,\mathrm{km\,s}^{-1}$). We thus find, assuming that the spectral line and the scattering molecular population's maxima align at the same frequency $\omega_0$, $\Gamma\simeq 1.7\times 10^{4}\,\mathrm{s}^{-1}$ and $L\simeq 1.8\times 10^6$~cm, while $N_g/g_g\simeq 1\times 10^{20}$. Incidentally, the value obtained for $\Gamma$, which we use at all frequencies, implies that the decay rate due to (non-coherent) collisions is negligible (i.e., $\gamma_\mathrm{coll}\sim 10^{-3}\,\mathrm{s}^{-1}$).

In order to use equation (\ref{eq:phi_p}) to calculate the Stokes parameters we also need to know the interaction time $t$. This will either be set by the physical size of the region through which the radiation propagates (and resonant scattering occurs) or the mean free path for the radiation, whichever is the shortest. An appropriate estimate for the mean free path $l_\mathrm{mp}$ is given by the inverse of the absorption coefficient (i.e., equation (\ref{eq:abs_coef})), and given the figures above we have $l_\mathrm{mp} = \alpha\left(\omega\right)^{-1}\sim 10^{16}$~cm and therefore $t\lesssim 10^6$~s.

Although we do not aim to provide a fit to the data presented in Figures \ref{fig:OrionKL_cp} and \ref{fig:OrionKL_lp}, as this would require modelling through a level of characterisation of the source we do not possess, we can use the parameters discussed above to inquire if our ARS model can qualitatively account for these observational results. More precisely, we have computed the response of such a system in the presence an ambient magnetic field of 2~mG oriented at 60~deg from the light of sight (i.e., $\iota=120$~deg), $l_\mathrm{mp}=2.5\times10^{16}$~cm (note that the magnetic field strength and $l_\mathrm{mp}$ are degenerate, i.e., a change in one parameter can be compensated by the inverse change in the other) and subjected to a weak incident linearly polarised signal (i.e., $1$~K or approximately $0.8\%$ polarisation level at the peak) at different angles $\varphi_0$ relative to the projection of the magnetic field on the plane of the sky. Also, while we modelled the molecular population with a Gaussian distribution, we use a Lorentzian profile ($\mathrm{FWHM}=25\,\mathrm{km\,s}^{-1}$) for the incident Stokes $I$ to better mimic that of the observations presented in Figure \ref{fig:OrionKL_cp} (top panel). This line shape is purely phenomenological, and only chosen to better model the high velocity wings stemming from the bipolar outflow (see, for example, \citealt{Erickson1982} and \citealt{Wilson1986}). More specifically, it is important to note that different parts of the line profile originate in environments that are quite different in nature, unlike the case of a Gaussian profile with thermal line broadening where one can assume that the emitting gas is co-spatial across the profile.

\begin{figure}
\begin{center}\includegraphics[trim=10 0 0 0, clip, scale=0.34]{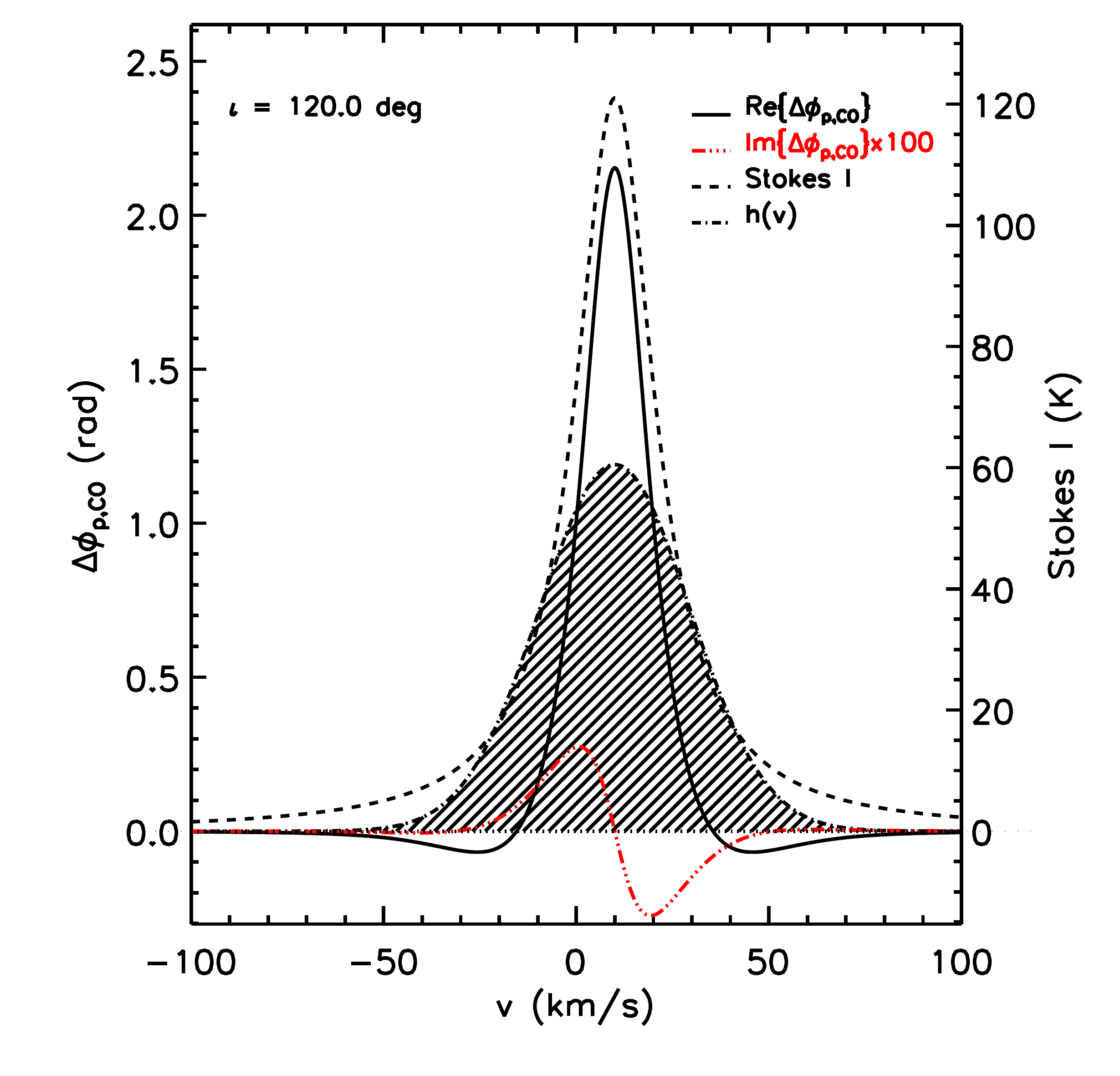}\end{center}
\caption{\label{fig:ARS_phi}Relative phase shift $\Delta\phi_{p,\mathrm{CO}}$ accrued between the two elliptical polarisation modes as a function of velocity (using the left vertical scale) specific to a system including a gas of CO molecules. The real component $\Delta\eta_{p,\mathrm{CO}}$ is shown with the solid black curve and the imaginary part $\Delta\gamma_{p,\mathrm{CO}}$ with the red triple-dot broken curve (times 100). The molecular distribution $h\left(v\right)$ is represented with the dot-dashed and hashed curve (normalised for display purposes) while the incident Lorentzian Stokes $I$ spectrum is shown using the dashed curve.}
\end{figure}

\begin{figure}
\begin{center}\includegraphics[trim=10 25 0 -25, clip, scale=0.34]{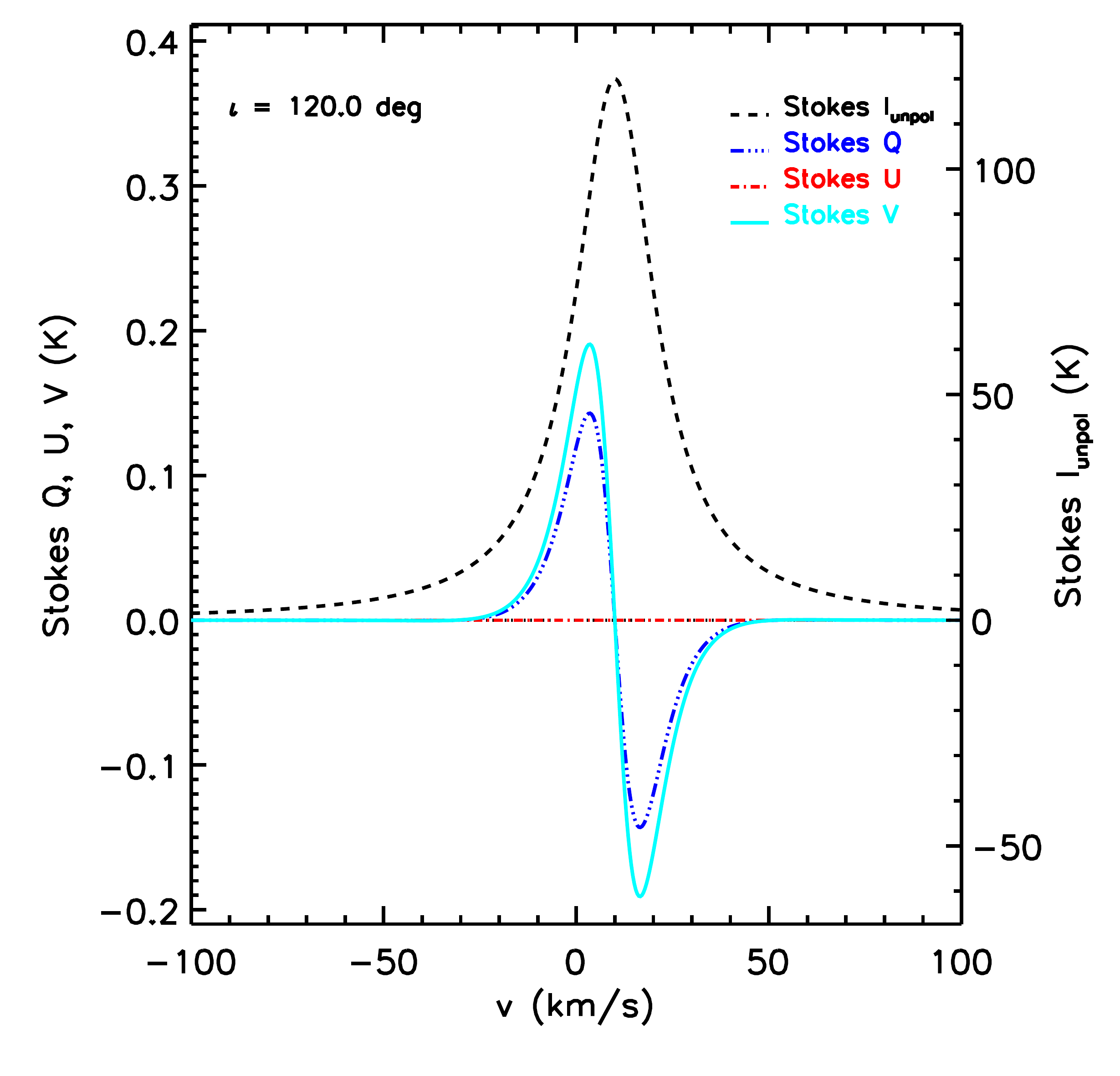}\end{center}
\caption{\label{fig:ARS_unpol}Polarisation signals generated from the unpolarised component of the incident signal (corresponding to the broken curve for Stokes $I_\mathrm{unpol}$ in the figure, right vertical scale) and is therefore independent of the angle $\varphi_0$ of the incident linear polarisation component. As expected from equations (\ref{eq:qp_unp})-(\ref{eq:vp_unp}), the linear polarisation can only appear in directions parallel or perpendicular to the orientation of the magnetic field on the plane of sky, i.e., Stokes $U$ (red dot-broken curve) is zero at all frequencies. The importance of the effect increases at smaller inclination angle $\iota$, while the relative strength between Stokes $Q$ (blue triple-dotted broken curve) and $V$ (cyan solid curve) is also a function of this parameter  (see equations (\ref{eq:qp_unp}-(\ref{eq:vp_unp})).}
\end{figure}

Figure \ref{fig:ARS_phi} shows the total relative phase shift $\Delta\phi_p$ accrued between the two elliptical polarisation modes $\Ket{n_{p,1}}$ and $\Ket{n_{p,2}}$ (see equations (\ref{eq:n_1})-(\ref{eq:n_2})) as a function of velocity (using the left vertical scale) specific to a system of CO molecules. The real component $\Delta\eta_p$ is shown with the solid black curve and the imaginary part $\Delta\gamma_p$ with the red triple-dot broken curve (times 100). The molecular distribution $h\left(v\right)$ is represented with the dot-dashed and hashed curve (normalised for display purposes) while the incident Lorentzian Stokes $I$ spectrum is shown using the dashed curve. The weaker term $\Delta\gamma_p$ is responsible for the generation of polarisation signals shown in Figure \ref{fig:ARS_unpol}. Since this component of the outgoing radiation stems from the unpolarised component of the incident signal (corresponding to the broken curve for Stokes $I_\mathrm{unpol}$ in the figure, using the vertical scale on the right), it is independent of the angle $\varphi_0$ of the incident linear polarisation component. In a manner consistent with equations (\ref{eq:qp_unp})-(\ref{eq:vp_unp}), it is seen from the figure that linear polarisation can only appear in directions parallel or perpendicular to the orientation of the magnetic field on the plane of sky (i.e, Stokes $U$ (red dot-broken curve) is zero at all frequencies). It is also observed that for the case treated here, where the peak of the molecular distribution is aligned with that of the incoming radiation, both the Stokes $Q$ (blue triple-dotted broken curve) and $V$ (cyan solid curve) profiles are anti-symmetric and reminiscent of what is expected for circular polarization stemming from the Zeeman effect. This specific behaviour disappears when the alignment between the centre of the profiles for the molecular distribution and the incoming radiation is removed. Likewise, the importance of the effect increases at smaller inclination angle $\iota$, while the relative strength between Stokes $Q$ and $V$ is also a function of this parameter (see equations (\ref{eq:qp_unp})-(\ref{eq:vp_unp})). 

\begin{figure}
\begin{center}\includegraphics[trim=10 0 0 0, clip, scale=0.34]{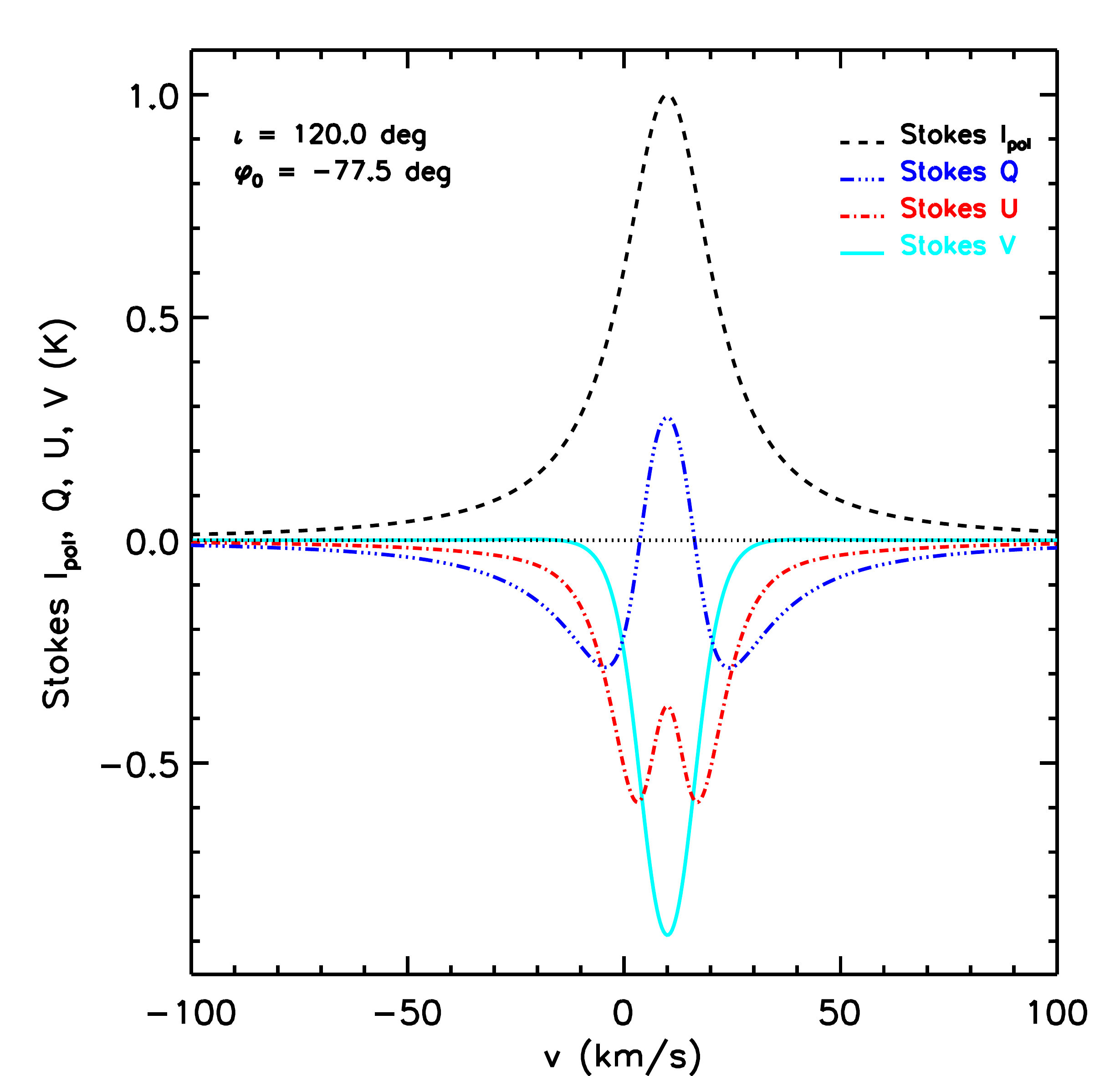}\end{center}
\caption{\label{fig:ARS_polarised}The Stokes  $Q$, $U$ and $V$ parameters resulting from the transformation of linearly polarised incident radiation of 1~K peak intensity, denoted by Stokes $I_\mathrm{pol}$ (broken curve). A significant amount of the incident linear polarisation is transformed into circular polarisation in the process, with the Stokes $V$ signal being the strongest nearing (minus) the intensity of the incident signal at the line centre.}
\end{figure}

The real component of the phase shift $\Delta\eta_p$ is responsible for the transformation of the incident linear polarisation signal. The case when $\varphi_0=-77.5$~deg for our example is shown in Figure \ref{fig:ARS_polarised}. Displayed on the figure are the Stokes $Q$, $U$ and $V$, as well as the polarised component of the incident radiation denoted by Stokes $I_\mathrm{pol}$ (broken curve). As can been seen a significant amount of the incident linear polarisation is transformed into circular polarisation in the process. In fact, at peak value the Stokes $V$ signal is the strongest at $\simeq 0.9$~K , nearing the maximum intensity of the incident polarised signal. 

\begin{center}
    \begin{figure*}
        \centering
        \includegraphics[trim=0 0 0 0, clip, scale=0.29]{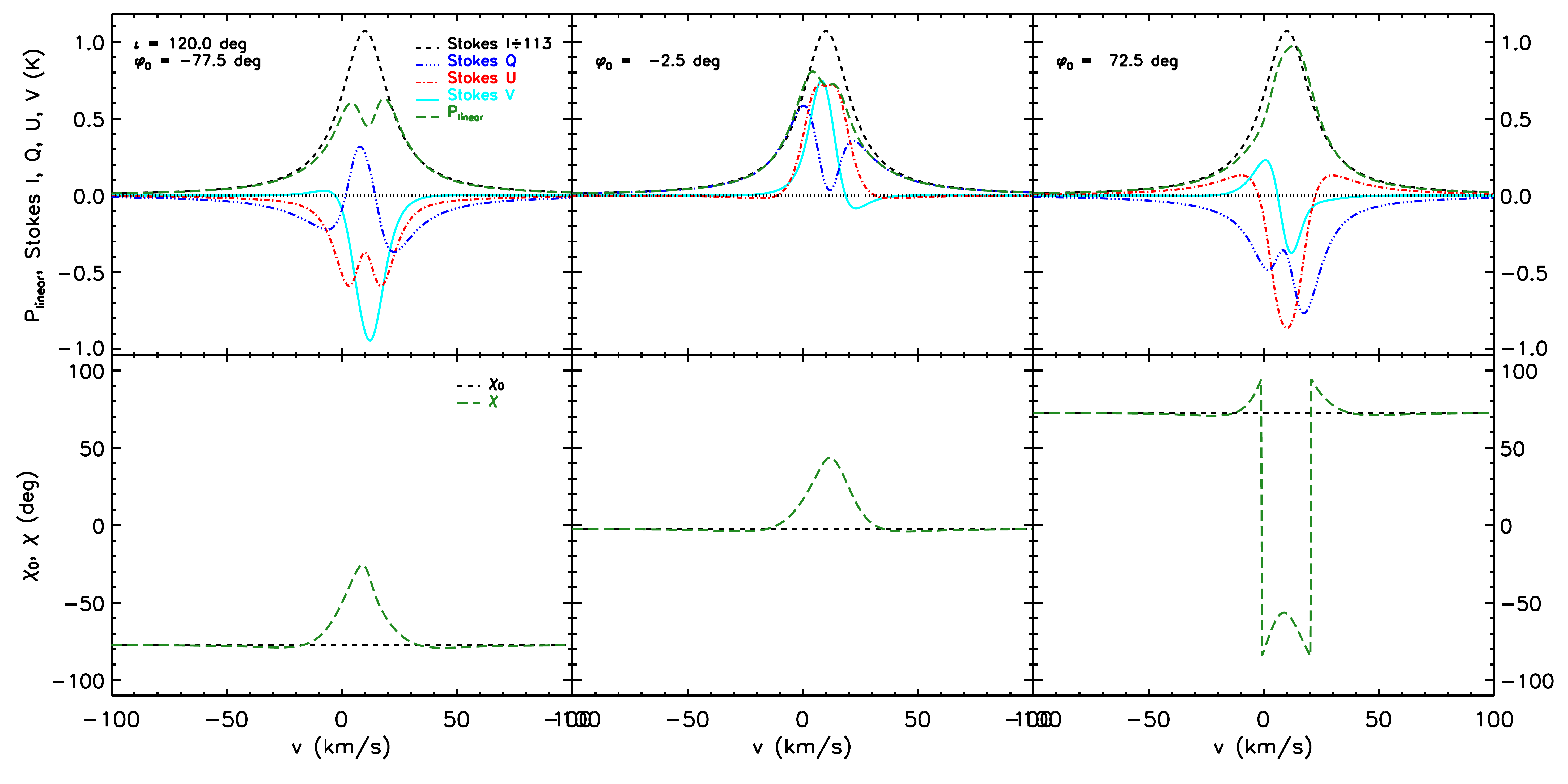}
        \caption{The total Stokes parameters consisting of the addition of the corresponding signals resulting from the unpolarised and polarised components of the incident radiation for three different values of incidence angle $\varphi_0=-77.5,\,-2.5$, and $72.5$~deg. In the top panel, the linear polarisation signal $P_\mathrm{linear}$ is shown (green broken curve) along with Stokes $I$, $Q$, $U$ and $V$, while in the bottom panels the polarisation angle of the linear polarisation for the incident ($\chi_0$, black broken curve) and scattered ($\chi$, green broken curve) radiation are displayed. The case for $\varphi_0=-77.5$~deg, which consists of the combination of the signals presented in Figures \ref{fig:ARS_unpol} and \ref{fig:ARS_polarised}, displays some of the same properties found in the profiles for the circular and linear polarisation signals measured in Orion KL and shown in Figures \ref{fig:OrionKL_cp} and \ref{fig:OrionKL_lp}, respectively.} 
        \label{fig:ARS_total}
    \end{figure*}
\end{center}

In Figure \ref{fig:ARS_total} we show the total Stokes parameters consisting of the combination of the corresponding signals resulting from the unpolarised (in Figure \ref{fig:ARS_unpol}) and polarised (as in Figure \ref{fig:ARS_polarised}) components of the incident radiation for three different values of incidence angle $\varphi_0=-77.5,\,-2.5$, and $72.5$~deg. The two components of a given Stokes parameter are combined in the manner explained in Appendix \ref{app:stokes} (see equation (\ref{eq:x_p_total_C})). In the top panels of the figure, the linear polarisation signal $P_\mathrm{linear}$ is shown (green broken curve) along with Stokes $I$, $Q$, $U$ and $V$, while in the bottom panels the polarisation angle of the linear polarisation for the incident ($\chi_0$, black broken curve) and scattered ($\chi$, green broken curve) radiation are displayed. More precisely, we have
\begin{align}
    & P_\mathrm{linear} = \sqrt{Q^2+U^2}\label{eq:P_linear}\\
    & \chi = \frac{1}{2}\arctan\left(\frac{U}{Q}\right).\label{eq:chi}
\end{align}

A comparison between the top three panels in Figure \ref{fig:ARS_total} gives an indication on the variety of profiles each Stokes parameters can take and their dependency on the incident polarisation angle $\chi_0=\varphi_0$. For example, the appearance of the Stokes $V$ line intensity can go from an almost symmetric profile (left panel at $\varphi_0=-77.5$~deg), to one showing the emergence of an inverted shoulder (centre panel at $\varphi_0=-2.5$~deg) and to a fully anti-symmetric shape reminiscent of the Zeeman effect (right panel at $\varphi_0=72.5$~deg). The case for $\varphi_0=-77.5$~deg, which consists of the combination of the signals presented in Figures \ref{fig:ARS_unpol} and \ref{fig:ARS_polarised}, displays some of the same properties found in the profiles for the circular and linear polarisation signals measured in Orion KL and shown in Figures \ref{fig:OrionKL_cp} and \ref{fig:OrionKL_lp}, respectively. In particular, we find that the Stokes $V$ spectrum has its peak slightly shifted to positive velocity compared to the Stokes $I$ profile, and $P_\mathrm{linear}$ displays a double-lobe feature with a slightly elevated right shoulder, similar to what is seen in the bottom panel of Figure \ref{fig:OrionKL_lp}. The difference in sign between our computed and observed Stokes $V$ spectra stems from the fact that \citet{Houde2013} used a different definition for this parameter \citep{Houde2014}. However, there are also differences such as in the width of the Stokes $V$ profile and a more pronounced change in the polarisation angle for the model of Figure \ref{fig:ARS_total}. We also find that the magnetic field strength (i.e., 2~mG) needed to recover the polarised intensities measured in Orion KL is as expected from previous measurements found in the literature \citep{Crutcher1999,Houde2009,Chuss2019}.

\section{Discussion and summary}\label{sec:summary}

We have developed in the previous sections a model for the interaction between a group of molecules, aligned under the influence of a local magnetic field, and an incident radiation field originating from some background source and propagating towards an observer far away in the foreground. The temporal evolution of this radiation/matter system happens through the fundamental processes of absorption, emission and scattering of photons. While our analysis originates from a starting point similar as that presented by \citet{Houde2013}, i.e., seeking to explain the discovery of circularly polarised signals in the CO$\left(J=2\rightarrow 1\right)$ rotational line in Orion KL through the ARS effect, our results quantitatively differ from theirs on some points. These differences stem from two key components in our new analysis, i.e., the realisation that \emph{i)} the resonance damping rate $\Gamma$ is not set by collisions but by the coherent radiative processes that determine the lifetimes of the virtual states involved in the scattering process, and \emph{ii)} the full solution to the problem must involve an elliptical polarisation basis (as opposed to a linear polarisation basis) to describe the eigenstates of the quantized radiation field at arbitrary magnetic field orientations. 

The determination of the resonance damping rate is an integral part of our solution to the problem and yielded a corresponding value that is orders of magnitude larger than previously thought for the Orion KL CO$\left(J=2\rightarrow 1\right)$ data. Accordingly, this implies significantly smaller correlation length $L=c/\Gamma$ and quantization volume $L^3$ for the radiation field, as well as a reduced number of molecules involved in the ARS process. The increased value for $\Gamma$ also brings into play the imaginary component $\Delta\gamma_p$ of the relative phase shift $\Delta\phi_p$($=\Delta\eta_p+i\Delta\gamma_p$) through which the ARS effect manifests itself on the polarisation state of the scattered radiation. More precisely, this  imaginary component $\Delta\gamma_p$ is responsible for the generation of polarised signals in an otherwise unpolarised incident radiation field (see Sec. \ref{sec:generation}). This effect was negligible in the analysis of \citet{Houde2013}.

While the use of a linear basis for expressing the radiation field initially revealed the existence of a Faraday conversion responsible for the transformation of linear polarisation into circular polarisation (and vice-versa), the elliptical polarisation basis further introduces a Faraday rotation of linearly polarised signals when the magnetic field is (anti-)aligned with the line of sight, and a mixture of the two effects at arbitrary inclination angles. The relative importance in the accrued phase shift $\Delta\phi_p$ of the term responsible for the Faraday rotation also largely compensated for the aforementioned reduced number of scattering molecules and preserved the strength of the ARS effect.  

Our calculations and application of the ARS effect to the Orion KL CO$\left(J=2\rightarrow 1\right)$ data of \citet{Houde2013} has confirmed that ARS is potentially unavoidable to explain the polarisation characteristics of some molecular rotational lines. First, the generation of polarisation signals is an inherent part of the effect. Although our example showed the levels of linear and circular polarisation expected from ARS to be on the order of a fraction of a percent for the case considered (see Fig. \ref{fig:ARS_unpol}), they could easily be significantly higher with slight variations on some parameters (e.g., smaller inclination angle, higher gas density, stronger magnetic field, lower excitation temperature, etc.) and brought to levels similar to those expected for the Goldreich-Kylafis effect (for linear polarisation; \citealt{Goldreich1981}). While ARS exhibits the same 90~deg ambiguity as the Goldreich-Kylafis effect for the orientation of the generated linear polarisation relative to that of the plane of the sky component of the magnetic field, it also displays further ambiguities for the inclination angle (`$\iota$' vs. `$\pi-\iota$') and the strength of the magnetic field (through the sign of $\Delta\gamma_p$; see Sec. \ref{sec:generation}). Second, ARS is highly efficient in transforming the polarisation state of background radiation signals through the coupling of the Stokes parameters (Secs. \ref{sec:conversion}, \ref{sec:rotation} and \ref{sec:transformation}). We have considered the case of incident linearly polarised signal for the Orion KL CO$\left(J=2\rightarrow 1\right)$ data, which could stem from either the Goldreich-Kylafis or the ARS effects in the background radiation source, and saw that, as mentioned earlier, a mixture of both Faraday conversion and Faraday rotation was responsible for the appearance of elliptical polarisation in the scattered radiation. This transformation property of the ARS is not limited to linear polarisation signals, however, and applies more generally to incident elliptical polarisation states.

It is important to note that the appearance of ARS polarisation signals in molecular rotational lines is entirely due to the Zeeman broadening of the line profiles. Namely, it is the slight relative displacements in frequency of the $\sigma_\pm$ spectral components between themselves and against the $\pi$ transition that are the sources of the relative phase shift $\Delta\phi_p$. Accordingly, and as we saw in Sec. \ref{sec:ARS}, the ARS effect disappears whenever $\omega_z\rightarrow 0$, i.e., as the magnetic field strength vanishes (see equations (\ref{eq:q_0})-(\ref{eq:v_0})). More precisely, while the Zeeman effect commonly discussed in the literature is for emission (or absorption) processes \citep{Crutcher2012} the effect underlying our analysis can, in fact, be directly attributed to the Zeeman effect for scattering. Although it is generally expected that scattering rates will be negligible compared to those due to emission and absorption processes for a single molecule at the wavelengths discussed here, the collective effect and the resulting coherence between molecules that characterise ARS greatly enhance the probability amplitude associated with forward scattering and renders the corresponding Zeeman effect (i.e., for scattering) observable for molecules such as CO (see below). These molecules/transitions are otherwise unaffected by the usual Zeeman effect in emission (or absorption). We note that the ARS enhancement is not limited to the Zeeman effect for scattering but would also apply to other phenomena responsible for the splitting of energy levels (e.g., the Stark effect). For that reason, and to avoid potential confusion, we will keep referring to the Zeeman effect as that which applies to emission or absorption processes and to ARS for the effect analysed and discussed in this paper.   

Given the relationship between the two phenomena, it should not be surprising that ARS also displays other characteristics usually associated with the Zeeman effect. For example, it is well established that Zeeman Stokes $V$ signals, which stem from the subtraction of intensities due to the $\sigma_\pm$ transitions, scale with the frequency derivative of the Stokes $I$ spectrum for a small Zeeman broadening \citep{Crutcher1993}. For ARS it is the relative phase shift $\Delta\phi_{p,\mathrm{c}}$ that results from the subtraction of corresponding components for the $\sigma_\pm$ transitions (see equations (\ref{eq:I_+-}) and (\ref{eq:Dphi_p,c})), and therefore scales with the frequency derivative of the scattering population distribution (i.e., in equation (\ref{eq:Ic(w)_app})) for the generation of polarisation signals. These derivative-like terms (for both the Zeeman and ARS effects) are proportional to the strength of the line of sight component of the magnetic field (see again equation (\ref{eq:Ic(w)_app})). In a similar manner, the Zeeman Stokes $Q$ and $U$ parameters for linear polarisation are proportional to the second order frequency derivative of the Stokes $I$ signal and the square of the strength of the plane of the sky component of the magnetic field. Because of their respective functionalities, it is found that $\left|V\right|\gg\left|Q\right|,\left|U\right|$ \citep{Crutcher1993}. Once again ARS somewhat mimics this behaviour in that it is the relative phase shift $\Delta\phi_{p,\mathrm{s}}$ that scales with the second-order frequency derivative of $h\left(\omega_\alpha\right)$ (see equations (\ref{eq:Dphi_p,s}) and (\ref{eq:Is(w)_app})) and the square of the strength of the plane of the sky component of the magnetic field (see again equation (\ref{eq:Is(w)_app})), while it is found that $\Delta\phi_{p,\mathrm{c}}\gg\Delta\phi_{p,\mathrm{s}}$. The similarities go no further, however, since for ARS the term due to the first-order frequency derivative of $h\left(\omega_\alpha\right)$ is also responsible for the generation of linear polarisation signals since the total relative phase shift $\Delta\phi_{p}\simeq\Delta\phi_{p,\mathrm{c}}$ enters in the equation for all Stokes parameters (see equations (\ref{eq:Stokes_q_pol})-(\ref{eq:Stokes_v_pol})). Also, as we have established in Sec. \ref{sec:ARS} and mentioned above, ARS is not only responsible for the generation of elliptically polarised signals from unpolarised background radiation but also for the transformation of the polarisation state of an incident electromagnetic field.  

Another fundamental difference between the ARS and Zeeman effects concerns the way they establish themselves within a molecular population. That is, any individual molecule sensitive to the Zeeman effect (e.g., CN) within a molecular gas subjected to an ambient magnetic field will experience a Zeeman broadening in its spectral transitions because of the influence of the Zeeman interaction term on the molecular energy levels. In that sense, the Zeeman effect is felt locally by all molecules that form the gas and can be observed as long as the magnetic field is strong enough, the molecular column density sufficient to build a detectable Stokes $V$ spectrum and velocity gradients or magnetic field reversals are not too pronounced to weaken or cancel Zeeman features along corresponding lines of sight. On the other hand, ARS is a collective effect between the constituents and cannot be described as happening on any given molecule with the effect increasing incrementally as the radiation scatters on individual molecules while propagating through the gas. In other words, ARS is not local. Rather, ARS will become detectable whenever a large enough number of molecules is present within the correlation volume characterising the radiation field ($L^3$ in our analysis). These molecules then act as a diffraction unit that enforces forward scattering as the dominant process and, as a consequence, renders ARS coherent in nature. This coherent behaviour from the group of molecules involved in the scattering of the background radiation manifests itself through the scaling of the relative phase shift $\Delta\phi_{p}$ with the number of molecules. 

Finally, given the predicted strength of ARS for molecules weakly sensitive to the Zeeman effect, such as CO, one may inquire about its potential importance for other molecules (e.g., CN) used to measure magnetic field strengths through the Zeeman effect. The Zeeman broadening for these transitions will be about three orders of magnitude larger than those for CO. One may therefore wonder if the polarisation signals measured with these molecules contain contributions from both the Zeeman effect and ARS, which could result in a significant complication for the interpretation of the corresponding measurements. Such instances could only be examined on a case by case basis as ARS will only become relevant when $\Delta\phi_{p}\sim 1$, a relation which is strongly dependent on the radiation coherence length $L$. A precise determination of this length scale would then become essential. 


These considerations and the analysis presented in this paper testify to the fact that the field of research focused on the study of magnetic fields through the polarisation characteristics of molecular rotational lines is becoming increasingly rich in its contents and its reach. Disentangling the effects from the different processes potentially at play, e.g., the Zeeman effect \citep{Crutcher1993,Crutcher1999}, the Goldreich-Kylafis effect \citep{Goldreich1981}, collisional polarisation \citep{Lankhaar2020} and ARS (\citealt{Houde2013} and this work), will necessitate all the tools at our disposal. As was exemplified by the work of \citet{Hezareh2013}, which we discussed in Sec. \ref{sec:Introduction}, comparisons with dust polarisation measurements will be essential. And most importantly, observations aimed at these processes should include spectra of all Stokes parameters as both linear and circular polarisation signals will hold clues vital for the successful completion of these studies.

\section*{Acknowledgements}
This paper is dedicated to the memory of R. H. Hildebrand. We thank the referee, H. Wiesemeyer, for his comments and constructive review. M.H.'s research is funded through the Natural Sciences and Engineering Research Council of Canada Discovery Grant RGPIN-2016-04460. M.H. is grateful for the hospitality of Perimeter Institute where part of this work was carried out. F.R.'s research at Perimeter Institute is supported in part by the Government of Canada through the Department of Innovation, Science and Economic Development Canada and by the Province of Ontario through the Ministry of Economic Development, Job Creation and Trade. B.L. acknowledges support from the Swedish Research Council (VR) through grant No. 2014-05713.

\section*{Data availability}
The data underlying this article will be shared on reasonable request to the corresponding author.

\bibliographystyle{mnras}
\bibliography{ARS-bib}

\appendix
\section{Global molecular state}\label{app:global_state}
The quantum mechanical state of a molecule will contain, in general, an internal component specifying levels of specific energies, as well as another for its external degrees of freedom (e.g., position, momentum, etc.). Although there are in principle several internal energy levels available to any molecule, we are interested in radiative transitions between sets of potentially degenerate lower $\left\{ \Ket{g_j}\right\}$ and excited $\left\{ \Ket{e_k}\right\}$ states of energies $E_{g}$ and $E_{e}$, respectively. If we assume that the molecule possesses a well defined linear momentum $\mathbf{p}$, then we can write its initial state as
\begin{equation}
    \Ket{\varphi_1} = \Ket{\mathbf{p}}\Ket{e_k}\label{eq:varphi_1}
\end{equation}
when in the excited state\footnote{More generally, the external and internal components in equation (\ref{eq:varphi_1}) can respectively be a superposition of different momenta and energy states. However, the simpler case considered here will be sufficient for our discussion.}. The electric-dipole interaction term of the Hamiltonian for this molecule can be written as
\begin{align}
    \hat{W}_1 = & -i\sum_{q}\mathcal{E}_{q}\hat{\mathbf{d}}\cdot\left[\boldsymbol{\epsilon}_{q}\hat{a}_{q}\hat{T}_\mathbf{p}\left(\hbar\mathbf{k}_{q}\right)-\boldsymbol{\epsilon}_{q}^*\hat{a}_{q}^{\dagger}\hat{T}_\mathbf{p}^\dagger\left(\hbar\mathbf{k}_{q}\right)\right],\label{eq:W_1}
\end{align}
where $\hat{T}_\mathbf{p}\left(\hbar\mathbf{k}_{q}\right)=e^{i\mathbf{k}_{q}\cdot\hat{\mathbf{r}}}$ is the linear momentum translation operator with $\hat{\mathbf{r}}$ the position operator for the molecule \citep{Haroche2008}. As in Sec. \ref{sec:theory}, $\mathcal{E}_{q}=\sqrt{\hbar\omega_{q}/2\epsilon_{0}L^{3}}$ is the one-photon electric field, with $L^{3}$ the quantization volume, $\hat{\mathbf{d}}$ the electric dipole moment of the molecule and $\hat{a}_{q}^{\dagger}$ ($\hat{a}_{q}$) the photon creation (annihilation) operator for the radiation mode $q$. The summation on $q$ is over the radiation  modes (wave vector and polarization) of polarisation states defined by $\mathbf{\boldsymbol{\epsilon}}_{q}$ and radiation frequency $\omega_{q}=\left|\mathbf{k}_q\right|c$ \citep{Grynberg2010}.

Acting on the state given by equation (\ref{eq:varphi_1}) with equation (\ref{eq:W_1}) can yield, for example, a transition to a new state
\begin{align}
    \Ket{\varphi_1^\prime} = \Ket{\mathbf{p}-\hbar\mathbf{k}_{q}}\Ket{g_j},\label{eq:varphi'_1} 
\end{align}
where the molecule was de-excited to the lower state $\Ket{g_j}$ and lost $\hbar\mathbf{k}_{q}$ in linear momentum through the emission of a photon of mode $\mathbf{k}_{q}$.

For our analysis, it will be convenient to work in the position basis $\Ket{\mathbf{r}}$. Correspondingly, the states of equations (\ref{eq:varphi_1}) and (\ref{eq:varphi'_1}) become
\begin{align}
    \Ket{\varphi_1} & = \int d^3r \ket{\mathbf{r}}\braket{\mathbf{r}|\varphi_1}\nonumber \\
    & = \frac{1}{\left(2\pi\hbar\right)^{3/2}} \int d^3r e^{i\mathbf{p}\cdot\mathbf{r}/\hbar}\Ket{\mathbf{r}}\Ket{e_k}\label{eq:varphi_1(r)}
\end{align}
and
\begin{align}
    \Ket{\varphi_1^\prime} & = \frac{1}{\left(2\pi\hbar\right)^{3/2}} \int d^3r e^{i\left(\mathbf{p}-\hbar\mathbf{k}_{q}\right)\cdot\mathbf{r}/\hbar}\Ket{\mathbf{r}}\Ket{g_j}.\label{eq:varphi'_1(r)}
\end{align}
If we now consider an ensemble of $N$ molecules of given initial distributions in linear momenta and internal states, then their global molecular state can be written as
\begin{align}
    \Ket{\varphi_N} = \frac{1}{\left(2\pi\hbar\right)^{3N/2}} \left[\prod_{\alpha=1}^N \int d^3r_\alpha e^{i\mathbf{p}_\alpha\cdot\mathbf{r}_\alpha/\hbar}\Ket{\mathbf{r}_\alpha}\right]\Ket{m},\label{eq:varphi(r)}
\end{align}
where $\Ket{m}$ is a global internal state formed by the direct product of the corresponding individual states (i.e., $\Ket{g_j}$, $\Ket{e_k}$, etc.) of each molecule (as defined in Sec. \ref{sec:theory}). It follows that the state of the system of $N$ molecules consists of a superposition of products of the type $\Ket{\mathbf{r}_1,\dots,\mathbf{r}_N}\Ket{m}$ that, because of the linearity of the Schr\"{o}dinger equation, can be treated independently. Each of these product state includes a phase term $e^{\sum_{\alpha=1}^N i\mathbf{p}_\alpha\cdot\mathbf{r}_\alpha/\hbar}$, which can be omitted while solving the Schr\"{o}dinger equation (as long as they are once again included afterwards if one wish to return to the linear momentum basis $\Ket{\mathbf{p}}$, for example). We then note that the only ``role'' of the position basis state $\Ket{\mathbf{r}_1,\dots,\mathbf{r}_N}$ in the process is to substitute $\hat{\mathbf{r}}\rightarrow\mathbf{r}$ in the interaction term of the Hamiltonian (compare equations (\ref{eq:W_1}) and (\ref{eq:W})), and introduce a term proportional to $e^{\pm i\mathbf{k}_q\cdot\mathbf{r}_\alpha}$ whenever a photon is absorbed or emitted (as in equation (\ref{eq:varphi'_1(r)})). Taking these considerations into account we will not explicitly write the position component $e^{\sum_{\alpha=1}^N i\mathbf{p}_\alpha\cdot\mathbf{r}_\alpha/\hbar}\Ket{\mathbf{r}_1,\dots,\mathbf{r}_N}$ in the global molecular state of the system in our analysis, with the understanding that it could be reinstated whenever needed.  

\section{Absorption/emission}\label{app:absorption}
\subsection{Probability amplitude}
From equation (\ref{eq:|psi'>_abs}), the first-order state of the system, with the terms for cases where transitions into and out of a state $\Ket{u_p}$ are separated, can be written as 
\begin{align}
\Ket{\psi^{\left(1\right)}} = & \,e^{-iE_{p}^{0}t/\hbar}\left[c_{p}-\frac{it}{\hbar}\sum_{n}c_{n}W_{pn}\right.\nonumber\\
 & \qquad\qquad\quad\left.\rule{0mm}{6mm}\times e^{i\frac{1}{2}\omega_{pn}t}\mathrm{sinc}\left(\frac{1}{2}\omega_{pn}t\right)\right]\Ket{u_{p}}\nonumber \\
 & -\frac{it}{\hbar}c_{p}\sum_{n\neq p}e^{-iE_{n}^{0}t/\hbar}W_{np}e^{i\frac{1}{2}\omega_{np}t}\mathrm{sinc}\left(\frac{1}{2}\omega_{np}t\right)\Ket{u_{n}}\nonumber \\
 & +\ldots,\label{eq:|psi'>_fo_A}
\end{align}
\noindent where $\Ket{u_{p}}=\Ket{m}\Ket{n_p}$ and $\Ket{u_{n}}=\Ket{m\pm1}\Ket{n_{n}}$, while the last line is meant to include all other terms present in $\Ket{\psi^{\left(1\right)}}$, which we will omit from now on. The quantum number $m$ is defined equation (\ref{eq:m}).

We first deal with the emission of one photon into state $\Ket{u_{p}}$ where
\begin{align}
W_{pn} = & \Braket{u_{p}|\hat{W}|u_{n}}\nonumber \\
 = & \,i\sum_{q}\sum_{\alpha=1}^{N_{e}+1}\,\mathcal{E}_{q}\left(\mathbf{\boldsymbol{\epsilon}}_{d}^*\cdot\boldsymbol{\epsilon}_{q}^*\right)e^{-i\mathbf{k}_{q}\cdot\mathbf{r}_{\alpha}}\braket{m|\hat{d}_{\alpha}|m+1}\braket{n_{p}|\hat{a}_{q}^{\dagger}|n_{n}}\nonumber \\
 = & \,i\sqrt{n_{r}+1}\,\mathcal{E}_{r}\left(\mathbf{\boldsymbol{\epsilon}}_{d}^*\cdot\boldsymbol{\epsilon}_{r}^*\right)d_\beta^* e^{-i\mathbf{k}_{r}\cdot\mathbf{r}_{\beta}},\label{eq:Wpn_abs_A}
\end{align}
\noindent since the two states can only differ in one atom (located at $\mathbf{r}_{\beta}$) and one radiation mode (mode $r$). If we now sum over all states $\Ket{u_{n}}$
\begin{align}
\sum_{n}c_{n}W_{pn}e^{i\frac{1}{2}\omega_{pn}t}\mathrm{sinc}\left(\frac{1}{2}\omega_{pn}t\right) = & \,id^* \sum_{r}\sum_{\beta=1}^{N_{e}+1}c_{r\beta}e^{-i\mathbf{k}_{r}\cdot\mathbf{r}_{\beta}}\nonumber \\
 & \hspace{-4cm} \times\sqrt{n_{r}+1}\,\mathcal{E}_{r}\left(\mathbf{\boldsymbol{\epsilon}}_{d}^*\cdot\boldsymbol{\epsilon}_{r}^*
 \right)e^{-i\frac{1}{2}\omega_{\beta r}t}\mathrm{sinc}\left(\frac{1}{2}\omega_{\beta r}t\right),\label{eq:sum_n_A}
\end{align}
\noindent with
\begin{align}
    \omega_{\beta r} & = \frac{p_\beta^2}{2M\hbar}+\omega_\beta-\frac{\left(\mathbf{p}_\beta-\hbar\mathbf{k}_r\right)^2}{2M\hbar}-\omega_r\nonumber\\
    & \simeq \omega_{\beta}-\omega_{r}\left(1-\frac{\mathbf{v}_\beta\cdot\mathbf{e}_{r}}{c}\right)\label{eq:omega_beta-r}
\end{align}
\noindent where $M$ is the molecular mass, $\mathbf{p}_\beta=M\mathbf{v}_\beta$ the linear momentum of molecule $\beta$ before the emission of the photon, $\mathbf{e}_r=\mathbf{k}_r/k_r$ and we assumed that all molecules have the same dipole moment $d$. We also substituted $\sum_{n}c_{n}\rightarrow \sum_{r}\sum_ {\beta}c_{r\beta}$ since the only states $\Ket{u_{n}}$ bringing a non-zero probability amplitude are those that can only differ for one atom (located at $\mathbf{r}_{\beta}$) and one radiation mode (mode $r$), as previously indicated. The number of molecules in an excited state $\Ket{e_k}$ in $\Ket{u_{n}}$ is given by $N_{e}+1$.

In all, there are a total of four different processes involving state $\Ket{u_p}=\Ket{m}\Ket{n_p}$: absorption of a photon from a molecular state $\Ket{m-1}$ and emission from a state $\Ket{m+1}$ into a state $\Ket{m}$; absorption into a state $\Ket{m+1}$ and emission into a state $\Ket{m-1}$ from a state $\Ket{m}$. We have dealt with the emission from a state $\Ket{m+1}$ above. Emission from a state $\Ket{m}$ yields the same result as in equation (\ref{eq:sum_n_A}) with the substitution $N_e+1\rightarrow N_e$. For the absorption of one photon we only need to consider the complex conjugate of equation (\ref{eq:sum_n_A}) while substituting $N_e+1\rightarrow N_{g}$ when out of a state $\Ket{m-1}$, as well as $n_r+1\rightarrow n_r$ and $N_e+1\rightarrow N_{g}+1$ when out of a state $\Ket{m}$. We recover equation (\ref{eq:|psi'>_abs-1}) when combining these four cases.

\subsection{Probability of absorption and emission}

For the probability calculations we start by focusing on the emission process into $\Ket{u_p}=\Ket{m}\Ket{n_p}$ in equation (\ref{eq:|psi'>_abs-1}) while assuming all states to be equally probable with $c_{r\beta}=c_{p}=c_0$ such that, in this case, for the state $\Ket{u_{p}}$ 
\begin{align}
\Ket{\psi_{p}^{\left(1\right)}\left(t\right)} \simeq &  \,c_0 e^{-iE_{p}^{0}t/\hbar}\left[1+\frac{t}{\hbar}d^*\sum_{r}\sum_{\beta=1}^{N_{e}+1}e^{-i\mathbf{k}_{r}\cdot\mathbf{r}_{\beta}}\sqrt{n_{r}+1}\,\mathcal{E}_{r}\right.\nonumber \\ 
& \left.\rule{0mm}{7mm}\times \left(\mathbf{\boldsymbol{\epsilon}}_{d}^*\cdot\boldsymbol{\epsilon}_{r}^*\right)e^{-i\frac{1}{2}\omega_{\beta r}t}\mathrm{sinc}\left(\frac{1}{2}\omega_{\beta r}t\right)\right]\Ket{u_{p}},\label{eq:|psi'_p>-1_A}
\end{align}
\noindent where it is understood that the summation on $r$ is for the radiation modes contained in $\Ket{u_p}$.

However, as stated in Sec. \ref{sec:absorption}, for the states relevant to the analysis only radiation modes pointing in the direction to the observer will contribute to the probability of emission (or absorption; see Figure \ref{fig:ARS}). If we further limit the radiation modes in $\Ket{u_p}$ to a specific frequency $\omega_{q}$ and take the square of the norm of the first order term to get the probability of emission into $\Ket{u_{p}}$ from the state with global molecular state $\Ket{m+1}$\footnote{To be more precise we should square the norm of the sum of the emission and absorption processes into $\Ket{u_p}$ in equation (\ref{eq:|psi'>_abs-1}). But because one summation is on $N_e+1$ molecules in upper states and the other on $N_g+1$ molecules in lower states, the cross terms will only involve two molecules common to both summations and can thus be safely neglected since $N_g, N_e\gg 1$ (see equation (\ref{eq:sum_alpha-beta_A})).} we get
\begin{align}
    \mathcal{P}_{\mathrm{em}}^{m+1}\left(\omega_{q},t\right) \simeq & \left|c_0\right|^{2}\frac{t^{2}}{\hbar^{2}}\left|d\right|^{2}n_{q}\,\mathcal{E}_{q}^{2}\left|\mathbf{\boldsymbol{\epsilon}}_{d}\cdot\boldsymbol{\epsilon}_{q}\right|^2\nonumber \\
    & \times \sum_{\alpha,\beta=1}^{N_{e}+1}e^{i\mathbf{k}_{q}\cdot\left(\mathbf{r}_{\alpha}-\mathbf{r}_{\beta}\right)}e^{i\frac{1}{2}\omega_{\alpha q}t}e^{-i\frac{1}{2}\omega_{q\beta}t}\nonumber\\
    & \times \mathrm{sinc}\left(\frac{1}{2}\omega_{\alpha q}t\right)\mathrm{sinc}\left(\frac{1}{2}\omega_{\beta q}t\right),\label{eq:Pse_A}
\end{align}
\noindent where $n_q\simeq n_q+1$ stands for all photons of frequency $\omega_q$ for the modes detected by the observer.

Energy conservation implies that $e^{i\frac{1}{2}\omega_{\alpha q}t}\mathrm{sinc}\left(\frac{1}{2}\omega_{\alpha q}t\right)\simeq e^{-i\frac{1}{2}\omega_{\beta q}t}\mathrm{sinc}\left(\frac{1}{2}\omega_{\beta q}t\right)\simeq1$ such that we can safely consider the summations on the molecular population independently from the rest of the equation. We can write for the expected value of the summations
\begin{align}
    \left\langle \sum_{\alpha,\beta=1}^{N_{e}+1}e^{i\mathbf{k}_{q}\cdot\left(\mathbf{r}_{\alpha}-\mathbf{r}_{\beta}\right)}\right\rangle  = & \sum_{\alpha=1}^{N_{e}+1}+ \sum_{\alpha\neq\beta}\left\langle e^{i\mathbf{k}_{q}\cdot\left(\mathbf{r}_{\alpha}-\mathbf{r}_{\beta}\right)}\right\rangle \nonumber \\
    = & \frac{g_g}{g_e}\left(N_{e}+1\right)\int h\left(\omega_{\alpha}\right)d\omega_{\alpha}\label{eq:sum_alpha-beta_A}
\end{align}
\noindent since the second term on the first line consists of a summation over random phases, which vanishes, and where $h\left(\omega_{\alpha}\right)$ is the normalized probability distribution function of the molecular Bohr frequency. We note that, while there are $N_{e}+1$ molecules in an excited state $\Ket{e_k}$ in $\Ket{u_{n}}$, the number of possible transitions for a given radiation mode is $g_g$. This has for consequence that the probability associated to emission will scale as $\left(N_e+1\right)g_g/g_e$, as indicated in equation (\ref{eq:sum_alpha-beta_A}), although that associated with absorption out of $\Ket{u_p}$ scales as $N_g$ (see equation (\ref{eq:sum_alpha-beta})). We therefore write
\begin{align}
    \mathcal{P}_{\mathrm{em}}^{m+1} & \left(\omega_{q},t\right) \simeq \left|c_0\right|^{2}\frac{t^{2}}{\hbar^{2}}\frac{g_g}{g_e}\left(N_{e}+1\right)\left|d\right|^{2}n_{q}\mathcal{E}_{q}^{2}\left|\mathbf{\boldsymbol{\epsilon}}_{d}\cdot\boldsymbol{\epsilon}_{q}\right|^2\nonumber \\
    & \qquad\qquad\times\int h\left(\omega_{\alpha}\right)\mathrm{sinc}^{2}\left(\frac{1}{2}\omega_{\alpha q}t\right)d\omega_{\alpha}\nonumber \\
    \simeq & \left|c_0\right|^{2}\frac{2\pi t}{\hbar^{2}} \frac{g_g}{g_e}\left(N_{e}+1\right)\left|d\right|^{2}n_{q}\mathcal{E}_{q}^{2}h\left(\omega_{q}\right)\left|\mathbf{\boldsymbol{\epsilon}}_{d}\cdot\boldsymbol{\epsilon}_{q}\right|^2,\label{eq:P_se_A}
\end{align}
\noindent where we used $\mathrm{sinc}^{2}\left(\frac{1}{2}\omega_{\alpha q}t\right)d\omega_{\alpha}\simeq\left(2\pi/t\right)\delta\left(x_{\alpha q}\right)dx_{\alpha}$ with $\omega_{\alpha q}$ defined as in equation (\ref{eq:omega_beta-r}). We also made the approximation $\omega_{q}\simeq\omega_{q}\left(1-\mathbf{v}_\alpha\cdot\mathbf{e}_{q}/c\right)$ in the last line, since $v_\alpha\ll c$ and to simplify the notation. 

Equation (\ref{eq:P_se_A}) can be further transformed by introducing the energy density $u\left(\omega_{q}\right)$ at $\omega_q$ and the Einstein spontaneous emission coefficient $A_{eg}$ for the $\Ket{e_k}\rightarrow\Ket{g_j}$ transition at the systemic velocity (see equations (\ref{eq:u(omega_p)}) and (\ref{eq:Aeg})) such that 
\begin{align}
    \mathcal{P}_{\mathrm{em}}^{m+1}\left(\omega_{q},t\right) & \simeq\left|c_0\right|^{2}t \frac{g_g}{g_e}\left(N_{e}+1\right)\frac{3\pi^{2}c^{3}}{\hbar\omega_{0}^{3}}\nonumber\\
    & \qquad\times A_{eg}u\left(\omega_{q}\right)h\left(\omega_{q}\right)\left|\mathbf{\boldsymbol{\epsilon}}_{d}\cdot\boldsymbol{\epsilon}_{q}\right|^2.\label{eq:P_se-1_A}
\end{align}
\noindent The corresponding transition rate is
\begin{align}
    \Gamma_{\mathrm{em}}^{m+1}\left(\omega_{q}\right) \simeq \frac{g_g}{g_e}\left(N_{e}+1\right)\frac{3\pi^{2}c^{3}}{\hbar\omega_{0}^{3}}A_{eg}u\left(\omega_{q}\right)h\left(\omega_{q}\right)\left|\mathbf{\boldsymbol{\epsilon}}_{d}\cdot\boldsymbol{\epsilon}_{q}\right|^2.\label{eq:Gamma_se_b_A}
\end{align}

Results identical in form are obtained for the other three processes discussed in the previous section by substituting the corresponding numbers of molecules involved. In general we have
\begin{align}
    & \Gamma_{\mathrm{abs}}^j\left(\omega_{q}\right) \simeq \left(\frac{N}{2}-j\right)\frac{3\pi^{2}c^{3}}{\hbar\omega_{0}^{3}}A_{eg}u\left(\omega_{q}\right)h\left(\omega_{q}\right)\left|\mathbf{\boldsymbol{\epsilon}}_{d}\cdot\boldsymbol{\epsilon}_{q}\right|^2\label{eq:Gamma_abs_a_A}\\
    & \Gamma_{\mathrm{em}}^j\left(\omega_{q}\right)\simeq \frac{g_g}{g_e}\left(\frac{N}{2}+j\right)\frac{3\pi^{2}c^{3}}{\hbar\omega_{0}^{3}}A_{eg}u\left(\omega_{q}\right)h\left(\omega_{q}\right)\left|\mathbf{\boldsymbol{\epsilon}}_{d}\cdot\boldsymbol{\epsilon}_{q}\right|^2\label{eq:Gamma_se_a_A}
\end{align}
\noindent with $N=N_g+N_e$ and $j=m, m\pm1$ where $m$ is defined in equation (\ref{eq:m}). From these equations we can define the rates of losses and gains for the state $\Ket{u_p}=\Ket{m}\Ket{n_p}$ with
\begin{align}
    & \Gamma_{p}^{-}\left(\omega_{q}\right) = \Gamma_{\mathrm{abs}}^m\left(\omega_{q}\right)+\Gamma_{\mathrm{em}}^m\left(\omega_{q}\right)\label{eq:Gamma-a_A}\\
    & \Gamma_{p}^{+}\left(\omega_{q}\right) = \Gamma_{\mathrm{abs}}^{m-1}\left(\omega_{q}\right)+\Gamma_{\mathrm{em}}^{m+1}\left(\omega_{q}\right).\label{eq:Gamma+a_A}
\end{align}
\noindent It is interesting to note that these two rates practically cancel each other out since $N_g,N_e\gg1$.

\section{Resonant scattering}\label{app:scattering}

Solving equation (\ref{eq:c2(t)_again}) for the second-order probability amplitude requires us to first evaluate the term $\braket{u_j|\hat{U}\left(t^\prime,t^{\prime\prime}\right)|u_k}$ on its right-hand side. To do so, we adopt the following perturbation expansion for the evolution operator \citep{Cohen-Tannoudji2017}
\begin{align}
    \hat{U}\left(t^\prime,\right. & \!\!\!\left.t^{\prime\prime}\right) = \hat{U}_0\left(t^\prime,t^{\prime\prime}\right)+\frac{1}{i\hbar}\int_{t^{\prime\prime}}^{t^\prime} d\tau\,\hat{U}_0\left(t^\prime,\tau\right)\hat{W}\hat{U}_0\left(\tau,t^{\prime\prime}\right)\nonumber\\
    & -\frac{1}{\hbar^2}\int_{t^{\prime\prime}}^{t^\prime} d\tau\int_{t^{\prime\prime}}^{\tau}d\tau^{\prime}\hat{U}_0\left(t^\prime,\tau\right)\hat{W}\hat{U}_0\left(\tau,\tau^{\prime}\right)\hat{W}\hat{U}_0\left(\tau^\prime,t^{\prime\prime}\right)\nonumber\\
    & + \ldots\label{eq:Uperturb_B}
\end{align}

The first (zeroth-order) term on the right-hand side is readily solved to yield
\begin{equation}
    \braket{u_j|\hat{U}_0\left(t^\prime,t^{\prime\prime}\right)|u_k} = e^{-iE_k^0\left(t^\prime-t^{\prime\prime}\right)/\hbar}\delta_{jk},\label{eq:0th_order_B}
\end{equation}
\noindent while the second (first-order) term was calculated to be zero in Sec. \ref{sec:absorption} on account of the random positioning of the molecules in space (see equation (\ref{eq:sum_beta-random})). For the last term we first write
\begin{align}
    \bra{u_j}\hat{U}_0\left(t^\prime,\tau\right)\hat{W} & \hat{U}_0\left(\tau,\tau^{\prime}\right)\hat{W}\hat{U}_0\left(\tau^\prime,t^{\prime\prime}\right)\ket{u_k} = e^{-iE_j^0t^\prime/\hbar}e^{iE_k^0t^{\prime\prime}/\hbar}\nonumber\\
    & \times  \sum_n e^{-i\omega_{nj}\tau}e^{i\omega_{nk}\tau^{\prime}}W_{jn}W_{nk},\label{eq:big_braket_B}
\end{align}
\noindent still with $W_{jn}=\braket{u_j|\hat{W}|u_n}$ and $\omega_{nk}=\left(E_n^0-E_k^0\right)/\hbar$. To simplify our discussion we will assume that the molecular states associated with $\Ket{u_j}$ and $\Ket{u_k}$ are of the type $\Ket{m+1}$ (or $\Ket{m^\prime+1}$, where both have $N_g-1$ molecules in a lower state but with potentially different arrangements) and we focus on the specific case where a photon is emitted from the state $\Ket{u_k}$. We will consider all possible cases later on. We thus write
\begin{align}
    \sum_n e^{-i\omega_{nj}\tau} & e^{i\omega_{nk}\tau^{\prime}}  W_{jn}W_{nk} = \sum_{r,q}\mathcal{E}_{r}\mathcal{E}_{q}\left(\mathbf{\boldsymbol{\epsilon}}_{d}\cdot\boldsymbol{\epsilon}_{r}\right)\left(\mathbf{\boldsymbol{\epsilon}}_{d}^*\cdot\boldsymbol{\epsilon}_{q}^*\right)\nonumber\\
    & \!\!\!\!\!\!\!\!\times\sum_{\alpha,\beta}^{N_e+1}e^{i\mathbf{k}_{r}\cdot\mathbf{r}_{\beta}}e^{-i\mathbf{k}_{q}\cdot\mathbf{r}_{\alpha}}\sum_n e^{-i\omega_{nj}\tau}e^{i\omega_{nk}\tau^{\prime}}\nonumber\\
    & \!\!\!\!\!\!\!\!\times\braket{m+1|\hat{d}_\beta|m}\braket{m|\hat{d}_\alpha|m^{\prime}+1}\braket{n_{j}|\hat{a}_{r}|n_{n}}\braket{n_{n}|\hat{a}_{q}^{\dagger}|n_{k}},\label{eq:sum_m_B}
\end{align}
\noindent where the summation on $n$ now implies $\sum_{n}=\sum_{m}\sum_{n_{n}}$, i.e., with the first summation on molecular states of type $\Ket{m}$ (which must have $N_g$ molecules in lower states) and the second on radiation modes. In equation (\ref{eq:sum_m_B}) we again assume the dipole moments to be the same for all molecules. Given this, and as was discussed in Sec. \ref{sec:ARS}, the double summation on $\alpha$ and $\beta$ for the molecules can be factored out and its expected value will bring in the following term
\begin{align}
    \left<\sum_{\alpha,\beta}^{N_e+1}e^{i\mathbf{k}_{r}\cdot\mathbf{r}_{\beta}}e^{-i\mathbf{k}_{q}\cdot\mathbf{r}_{\alpha}}\right> & = \sum_{\alpha,\beta}^{N_e+1}\left<e^{i\mathbf{k}_{r}\cdot\mathbf{r}_{\beta}}e^{-i\mathbf{k}_{q}\cdot\mathbf{r}_{\alpha}}\right>\nonumber\\
    & \hspace{-20mm} = \sum_{\alpha}^{N_e+1}\left<e^{-i\left(\mathbf{k}_{q}-\mathbf{k}_{r}\right)\cdot\mathbf{r}_{\alpha}}\right>+\sum_{\beta\neq\alpha}^{N_e+1}\left<e^{i\mathbf{k}_{r}\cdot\mathbf{r}_{\beta}}\right>\left<e^{-i\mathbf{k}_{q}\cdot\mathbf{r}_{\alpha}}\right>\nonumber\\
    & \hspace{-20mm} =\sum_{\alpha}^{N_e+1}\left<e^{-i\left(\mathbf{k}_{q}-\mathbf{k}_{r}\right)\cdot\mathbf{r}_{\alpha}}\right>,\label{eq:sum_molecules_B}
\end{align}
\noindent which implies that the two transitions in equation (\ref{eq:sum_m_B}) must happen in only one molecule. The same will be true if a photon is absorbed from state $\Ket{u_k}$ instead of being emitted, or if the molecular states associated with $\Ket{u_j}$ and $\Ket{u_k}$ are of type $\Ket{m-1}$ instead of $\Ket{m+1}$. However, whenever $\Ket{u_j}$ and $\Ket{u_k}$ are associated to different types of molecular states (e.g., when $\Ket{u_j}=\Ket{m+1}\Ket{n_j}$ and $\Ket{u_k}=\Ket{m-1}\Ket{n_k}$, etc.) we must have a minimum of two molecules differing in their states (i.e., they have respectively have $N_g-1$ and $N_g+1$ molecules in lower states) and the condition that the two transitions happen for one molecule cannot be met. This will result in a vanishing probably amplitude in equation (\ref{eq:big_braket_B}). We therefore conclude that $\Ket{u_j}$ and $\Ket{u_k}$ must share the same type of molecular state, i.e., $\Ket{m+1}$ and $\Ket{m^\prime+1}$ for the case considered here. Furthermore, the only difference between these two states can only happen for the molecule hosting the transitions (e.g., that molecule can be in state $\Ket{e_k}$ for $\Ket{m+1}$ and $\Ket{e_{k^\prime}}$ for $\Ket{m^\prime+1}$ with $k\neq k^\prime$).

We can go further by transforming the exponential in equation (\ref{eq:sum_molecules_B}) using cylindrical coordinates and write 
\begin{equation}
\left<e^{-i\left(\mathbf{k}_{q}-\mathbf{k}_{r}\right)\cdot\mathbf{r}_{\alpha}}\right> = \left<e^{-i\left(k_{q}\cos\theta-k_{r}\right)z_\alpha}\right>\left<e^{-ik_{q}\rho_\alpha\sin\theta\cos\left(\phi-\phi_\alpha\right)}\right>\label{eq:average_break_B} 
\end{equation}
\noindent since the product of the two averages results from the fact that the positioning of the molecules along the different axes is statistically independent. We also set the orientation of $\mathbf{k}_{r}$ as a reference and expressed $\mathbf{k}_{q}$ in relation to it.

The summation on the molecules on the right-hand side of equation (\ref{eq:sum_molecules_B}) is nothing more than a diffraction relation and, given that the wavelength $\lambda\ll L$, we therefore expect $\theta\ll1$ in equation (\ref{eq:average_break_B}). Thus keeping the exponent to first order in $\theta$ in the first average on the right-hand side we have
\begin{align}
\left\langle e^{-i\left(k_{q}\cos\theta-k_{r}\right)z_{\alpha}}\right\rangle  \simeq & \left\langle e^{-i\Delta k_{qr}z_{\alpha}}\right\rangle \nonumber \\
 \simeq & \frac{1}{L}\int_{0}^{L}e^{-i\Delta k_{qr}z_{\alpha}}dz_{\alpha}\nonumber \\
 \simeq & \,e^{-i\frac{1}{2}\Delta k_{qr}L}\mathrm{sinc}\left(\frac{1}{2}\Delta k_{qr}L\right)\label{eq:exp_z_B}
\end{align}
\noindent with $\Delta k_{qr}=k_q-k_r$ and where we assumed a uniform probability distribution for the position of a molecule. This relation is very strongly peaked at unity and sets the following constraint
\begin{equation}
\Delta k_{qr}\lesssim\frac{2\pi}{L},\label{eq:Dk_qr_B}
\end{equation}
\noindent which, given the expected size of the system, implies that we can safely set $k_{q}=k_{r}$ from now on. 

Turning our attention to the second average on the right-hand side in equation (\ref{eq:average_break_B}), we have
\begin{align}
\left<e^{-ik_{q}\rho_\alpha\sin\theta\cos\left(\phi-\phi_\alpha\right)}\right>  = & \frac{4}{\pi L^{2}}\int_{0}^{L/2}d\rho_\alpha\rho_\alpha\nonumber \\
& \times\int_{0}^{2\pi}d\phi_\alpha e^{-ik_{q} \rho_\alpha\sin\theta\cos\left(\phi-\phi_\alpha\right)}\nonumber \\
 = & \frac{2 J_{1}\left(\frac{1}{2}k_{q}L\sin\theta\right)}{\frac{1}{2}k_{q}L\sin\theta}.\label{eq:exp_xy_B}
\end{align}
\noindent The right-hand side of equation (\ref{eq:exp_xy_B}) is very strongly peaked in $\theta$, since $L\gg\lambda$, has a maximum of one at $\theta=0$ and first reaches zero at $k_{q}L\theta/2\simeq3.83$, which implies the following constraint
\begin{equation}
\theta\lesssim\frac{\lambda}{L}.\label{eq:theta_B}
\end{equation}
\noindent Equation (\ref{eq:theta_B}) justifies our previous assumption when dealing with equation (\ref{eq:exp_z_B}). The constraints given by equations (\ref{eq:Dk_qr_B}) and (\ref{eq:theta_B}) imply that $\mathbf{k}_{q}=\mathbf{k}_{r}$.

Most importantly, the combination of the condition of having the same type of globular molecular state (e.g., $\Ket{m+1}$ and $\Ket{m^\prime+1}$) with $\mathbf{k}_{q}=\mathbf{k}_{r}$ also implies that $\Ket{u_j}=\Ket{u_k}$ in equation (\ref{eq:big_braket_B}) since the photon emitted from $\Ket{u_k}$ is removed through the absorption of another photon of similar characteristics into $\Ket{u_j}$ (different polarisation states will occur at different frequencies because of the Zeeman effect). We can then rewrite equation (\ref{eq:sum_m_B}) as
\begin{align}
    \sum_n e^{-i\omega_{nj}\tau} & e^{i\omega_{nk}\tau^{\prime}}W_{jn}W_{nk} = \left|d\right|^{2}\sum_{q}\left(n_q+1\right)\mathcal{E}_{q}^2\left|\mathbf{\boldsymbol{\epsilon}}_{d}\cdot\boldsymbol{\epsilon}_{q}\right|^2 \nonumber\\
    & \hspace{-5mm}\times \frac{g_g}{g_e}\left(N_e+1\right)\int d\omega_\alpha h\left(\omega_\alpha\right)e^{i\omega_{\alpha q} \left(\tau-\tau^\prime\right)}\delta_{jk},\label{eq:sum_m_2_B}
\end{align}
\noindent where as before $\omega_{\alpha q}=\omega_\alpha-\omega_q\left(1-\mathbf{v}_\alpha\cdot\mathbf{e}_{q}/c\right)$, and for equation (\ref{eq:big_braket_B}) 
\begin{align}
    \bra{u_j}\hat{U}_0\left(t^\prime,\tau\right)\hat{W} & \hat{U}_0\left(\tau,\tau^{\prime}\right)\hat{W}\hat{U}_0\left(\tau^\prime,t^{\prime\prime}\right)\ket{u_k} = \delta_{jk}e^{-iE_k^0\left(t^\prime-t^{\prime\prime}\right)/\hbar}\nonumber\\
    & \times \frac{g_g}{g_e}\left(N_e+1\right)\left|d\right|^{2}\sum_{q}\left(n_q+1\right)\mathcal{E}_{q}^2\left|\mathbf{\boldsymbol{\epsilon}}_{d}\cdot\boldsymbol{\epsilon}_{q}\right|^2\nonumber\\
    & \times\int d\omega_\alpha h\left(\omega_\alpha\right)e^{i\omega_{\alpha q}\left(\tau-\tau^\prime\right)}.\label{eq:big_braket_2_B}  
\end{align}

The insertion of this relation into equation (\ref{eq:Uperturb_B}) will bring the following integral
\begin{align}
    \int d & \omega_\alpha h\left(\omega_\alpha\right)\int_{t^{\prime\prime}}^{t^\prime} d\tau\int_{t^{\prime\prime}}^{\tau}d\tau^{\prime}e^{i\omega_{\alpha q}\left(\tau-\tau^\prime\right)}\nonumber\\ 
    & = \frac{1}{2}\int d\omega_\alpha h\left(\omega_\alpha\right)\int_{0}^{t^\prime-t^{\prime\prime}} d\lambda e^{i\omega_{\alpha q}\lambda}\int_{0}^{2\left(t^\prime-t^{\prime\prime}\right)} d\lambda^\prime\nonumber\\
    & \simeq \left(t^\prime-t^{\prime\prime}\right)\int d\omega_\alpha h\left(\omega_\alpha\right)\int_{0}^{\infty} d\lambda e^{i\omega_{\alpha q}\lambda}\nonumber\\
    & \simeq \left(t^\prime-t^{\prime\prime}\right)\int d\omega_\alpha h\left(\omega_\alpha\right)\left[\pi\delta\left(\omega_{\alpha q}\right)+i\mathcal{P}\frac{1}{\omega_{\alpha q}}\right]\nonumber\\
    & \simeq \left(t^\prime-t^{\prime\prime}\right)\left[\pi h\left(\omega_q\right)+i\int d\omega_\alpha\mathcal{P}\frac{h\left(\omega_\alpha\right)}{\omega_{\alpha q}}\right],\label{eq:Principal_part_B}
\end{align}
\noindent where we set $\lambda=\tau-\tau^\prime$ and $\lambda^\prime=\tau+\tau^\prime-2t^{\prime\prime}$, once again approximated $\omega_{q}\simeq\omega_{q}\left(1-\mathbf{v}_\alpha\cdot\mathbf{e}_{q}/c\right)$ in the last line, while $\mathcal{P}$ stands for Cauchy's principal value. In going from the second to the third line we took advantage of the fact that the (inverse) Fourier transform of $h\left(\omega_\alpha\right)$ is of limited temporal width (in $\lambda$-space) to extend the upper bound of the integral on $\lambda$ from $t^\prime-t^{\prime\prime}$ to infinity.

The same calculations can be performed for the absorption of a photon from state $\Ket{u_k}=\Ket{m+1}\Ket{n_k}$, and then again when $\Ket{u_k}=\Ket{m-1}\Ket{n_k}$. Combining all of these results, it can be shown that
\begin{align}
    \braket{u_j|\hat{U}\left(t^\prime,t^{\prime\prime}\right)|u_k} & \simeq \delta_{jk}e^{-iE_k^0\left(t^\prime-t^{\prime\prime}\right)/\hbar}\left[1-\left(\frac{\Gamma}{2}+i\frac{\Phi}{\hbar}\right)\left(t^\prime-t^{\prime\prime}\right)\right]\nonumber\\
    & \simeq \delta_{jk}e^{-i\left(E_k^0+\Phi\right)\left(t^\prime-t^{\prime\prime}\right)/\hbar}e^{-\frac{1}{2}\Gamma\left(t^\prime-t^{\prime\prime}\right)}\label{eq:<uj|U|uk>_B},
\end{align}
\noindent where we assumed $\Gamma,\,\Phi/\hbar\ll 1/\left(t^\prime-t^{\prime\prime}\right)$ in the last line and (see equations (\ref{eq:Gamma_abs_a_A})-(\ref{eq:Gamma_se_a_A}))
\begin{align}
    \Gamma & = \sum_r\left[\Gamma^{m+1}_\mathrm{em}\left(\omega_r\right)+\Gamma^{m+1}_\mathrm{abs}\left(\omega_r\right)+\Gamma^{m-1}_\mathrm{em}\left(\omega_r\right)+\Gamma^{m-1}_\mathrm{abs}\left(\omega_r\right)\right]\nonumber\\
    & \simeq 2\left(1+\frac{g_g}{g_e}\frac{N_e}{N_g}\right)\sum_r\Gamma^m_\mathrm{abs}\left(\omega_r\right)\label{eq:Gamma_B}\\
    \Phi & = -N_g\left(1-\frac{g_g}{g_e}\frac{N_e}{N_g}\right)\frac{3\pi c^{3}}{\omega_{0}^{3}}A_{eg}\sum_r u\left(\omega_{r}\right)\left|\mathbf{\boldsymbol{\epsilon}}_{d}\cdot\boldsymbol{\epsilon}_{r}\right|^2\nonumber\\
    & \quad\times \int d\omega_\alpha\mathcal{P}\frac{h\left(\omega_\alpha\right)}{\omega_{\alpha r}}.\label{eq:Phi_B}
\end{align}
\noindent We therefore see that $\Gamma$ is the sum of the radiation rates associated with all possible states for $\Ket{u_k}$, while $\Phi\left(\ll\hbar\omega_\alpha\right)$ is responsible for a shift in its unperturbed energy level $E_k^0$ \citep{Cohen-Tannoudji2018b}.  
  
We are now in a position to determine the probability amplitude for resonant scattering by inserting equation (\ref{eq:<uj|U|uk>_B}) into equation (\ref{eq:c2(t)_again})
\begin{align}
    c_{p}^{\left(2\right)} & \left(t\right) = -\frac{1}{\hbar^2}e^{-iE_{p}^{0}t/\hbar}\sum_{n}\sum_{k}c_n W_{pk}W_{kn}\nonumber\\
    & \times\int_0^t dt^\prime e^{i\omega_{pk}^\prime t^\prime}e^{-\frac{1}{2}\Gamma t^\prime}
    \int_0^{t^\prime}dt^{\prime\prime}e^{-i\omega_{nk}^\prime t^{\prime\prime}}e^{\frac{1}{2}\Gamma t^{\prime\prime}}\nonumber\\
    & = \frac{it}{\hbar^2}e^{-iE_{p}^{0}t/\hbar}\sum_{n}\sum_{k}c_n\frac{W_{pk}W_{kn}}{\omega_{kn}^+-i\Gamma/2}e^{i\frac{1}{2}\omega_{pn}t}\mathrm{sinc}\left(\frac{1}{2}\omega_{pn}t\right),\label{eq:c2(t)_3_B}
\end{align}
\noindent where $\omega_{kn}^+=\omega_{kn}+\Phi/\hbar$ and we only kept the term that preserves energy conservation between in the initial and final states when solving the integrals \citep{Schiff1968}. We therefore see that the main effect of the displacement in energy $\Phi$ is to bring a shift in the Bohr frequency of transitions between states $\Ket{u_n}$ and $\Ket{u_k}$ \citep{Cohen-Tannoudji2017}. Otherwise, this energy displacement will not have any incidence on the outcome of our analysis. 

Solving equation (\ref{eq:c2(t)_3_B}) for the probability amplitude of resonant scattering where $\Ket{u_p}=\Ket{m}\Ket{n_p}$ and $\Ket{u_n}=\Ket{m^\prime}\Ket{n_n}$ hinges on handling the summation on the intermediate states $\Ket{u_k}$. We will proceed in a manner similar to what was done previously for equation (\ref{eq:big_braket_B}). That is, first focusing on the absorption of an incident photon, at the heart of our calculations is the term
\begin{align}
\sum_{k}\frac{W_{pk}W_{kn}}{\omega_{kn}^+-i\Gamma/2} & = \sum_{k}\sum_{r,q}\mathcal{E}_{r}\mathcal{E}_{q}\left(\mathbf{\boldsymbol{\epsilon}}_{d}^*\cdot\boldsymbol{\epsilon}_{r}^*\right)\left(\mathbf{\boldsymbol{\epsilon}}_{d}\cdot\boldsymbol{\epsilon}_{q}\right)\nonumber \\
 & \hspace{-10mm}\times\sum_{\beta,\alpha}^{N_g}\frac{e^{-i\mathbf{k}_{r}\cdot\mathbf{r}_{\beta}}e^{i\mathbf{k}_{q}\cdot\mathbf{r}_{\alpha}}}{\omega_\alpha^+-\omega_q-i\Gamma/2}\nonumber \\
 & \hspace{-10mm}\times\braket{m|\hat{d}_\beta|m+1}\braket{m+1|\hat{d}_\alpha|m^{\prime}}\braket{n_{p}|\hat{a}_{r}^{\dagger}|n_{k}}\braket{n_{k}|\hat{a}_{q}|n_{n}}, \label{eq:sum_j_B}
\end{align}
\noindent where the summation on the virtual states implies $\sum_{k}=\sum_{m+1}\sum_{n_k}$, i.e., with the first summation on molecular states and the second on radiation modes, and $\omega_\alpha^\pm=\omega_\alpha\pm\Phi/\hbar$ and we approximated $\omega_{q}\simeq\omega_{q}\left(1-\mathbf{v}_\alpha\cdot\mathbf{e}_{q}/c\right)$.

Equation (\ref{eq:sum_j_B}) is in many ways similar to equation (\ref{eq:sum_m_B}). In particular the double summation on the molecules can be dealt as previously with
\begin{align}
    \left<\sum_{\alpha,\beta}^{N_g}\frac{e^{-i\mathbf{k}_{r}\cdot\mathbf{r}_{\beta}}e^{i\mathbf{k}_{q}\cdot\mathbf{r}_{\alpha}}}{\omega_\alpha^+-\omega_q-i\Gamma/2}\right> & = \sum_{\alpha,\beta}^{N_g}\frac{\left<e^{-i\mathbf{k}_{r}\cdot\mathbf{r}_{\beta}}e^{i\mathbf{k}_{q}\cdot\mathbf{r}_{\alpha}}\right>}{\omega_\alpha^+-\omega_q-i\Gamma/2}\nonumber\\
    & = \sum_{\alpha}^{N_g}\frac{\left<e^{-i\left(\mathbf{k}_{q}-\mathbf{k}_{r}\right)\cdot\mathbf{r}_{\alpha}}\right>}{\omega_\alpha^+-\omega_q-i\Gamma/2},\label{eq:average_B} 
\end{align}
\noindent where equation (\ref{eq:sum_molecules_B}) was used. We thus find that the scattering can only involve one molecule. Following the procedure laid out in equations (\ref{eq:average_break_B})-(\ref{eq:theta_B}) we have
\begin{equation}
    \left<\sum_{\alpha,\beta}^{N_g}\frac{e^{-i\mathbf{k}_{r}\cdot\mathbf{r}_{\beta}}e^{i\mathbf{k}_{q}\cdot\mathbf{r}_{\alpha}}}{\omega_\alpha^+-\omega_q-i\Gamma/2}\right> = N_g\int\frac{h\left(\omega_\alpha\right)d\omega_\alpha}{\omega_\alpha^+-\omega_q-i\Gamma/2}.\label{eq:sum_molecules_2_B}
\end{equation}

A similar term exists for the emission of a photon into the virtual state $\Ket{u_k}=\Ket{m-1}\Ket{n_k}$. Assuming $n_q+1\simeq n_q$, we simply have to replace $N_g$ by $N_e g_g/g_e$, and substitute $\omega_\alpha^+\rightarrow -\omega_\alpha^-$ and $\omega_q\rightarrow -\omega_q$ in equation (\ref{eq:sum_molecules_2_B}) to obtain 
\begin{align}
    c_{p}^{\left(2\right)}\left(t\right) & = c_p\frac{it}{\hbar}e^{-iE_{p}^{0}t/\hbar}\frac{3\pi c^{3}}{2\omega_{0}^{3}}A_{eg}\sum_{q}u\left(\omega_q\right)\left|\mathbf{\boldsymbol{\epsilon}}_{d}\cdot\boldsymbol{\epsilon}_{q}\right|^{2}\nonumber\\
    & \hspace{-2mm}\times \left[N_g\int\frac{h\left(\omega_\alpha\right)d\omega_\alpha}{\omega_\alpha^+-\omega_q-i\Gamma/2}-\frac{g_g}{g_e}N_e\int\frac{h\left(\omega_\alpha\right)d\omega_\alpha}{\omega_\alpha^- -\omega_q+i\Gamma/2}\right]\label{eq:c2(t)_4_B}
\end{align}
\noindent since it must also be the case that $\Ket{u_n}=\Ket{u_p}$ (see the discussion following equation (\ref{eq:theta_B})). Finally, inserting equation (\ref{eq:c2(t)_4_B}) into equation (\ref{eq:|psi(t)_exp}) (with $c_p^{\left(1\right)}=0$) yields equation (\ref{eq:|psi'_p>-2}).

\section{Elliptical basis}\label{app:elliptical}

Our solution to the problem of the interaction between a large number of molecules and an incident radiation field established the dominance of the forward resonant scattering process. One of the consequences of this is the realisation that compound eigenstates (i.e., molecular and radiation together) for the non-interacting Hamiltonian $\hat{H}_0$ are also eigenstates for the interacting problem\footnote{This can be verified by replacing the interaction Hamiltonian $\hat{W}$ by an effective interaction Hamiltonian defined by the second order interaction term, which is proportional to $\left|d\right|^2$ \citep{Grynberg2010}. This is feasible because the probability amplitude of the interaction Haniltonian's first order term vanishes (see Sec. \ref{sec:absorption} and Appendix \ref{app:absorption}). This effective interaction Hamiltonian will have the same symmetry as $\hat{H}_0$, and therefore share its eigenstates, since $\left|d\right|^2$ is totally symmetric under the molecular symmetry (MS) group of a single molecule \citep{Bunker1998}.}. It therefore follows that the molecular states of type $\Ket{m}$ (as defined in the discussion surrounding equation (\ref{eq:m})), and their linear combinations, used in the analysis are the corresponding eigenstates for the matter component. However, the radiation eigenstates for a given polarisation and mode (i.e., $\Ket{n_{p,j}}$ with $j=1,2$ in Sec. \ref{sec:ARS}) were assumed to exist and never explicitly written down. In this appendix we show how this basis can be determined from symmetry considerations. 

We first note that $\hat{H}_0$ contains the Zeeman interaction term responsible for the splitting of the $\pi$ and $\sigma_\pm$ spectral components. Because the magnetic field is assumed responsible for the alignment of the molecules' electric dipole moment, it will set the rotational symmetry of the Hamiltonian. More precisely, we can express the electric dipole moment using the basis characterizing the $\pi$ and $\sigma_\pm$ spectral transitions through \citep{Cohen-Tannoudji2018b}
\begin{align}
    \hat{\mathbf{d}} & = \hat{d}_z\mathbf{e}_{z}+\hat{d}_x\mathbf{e}_{x}+\hat{d}_y\mathbf{e}_{y}\nonumber\\
                     & = \hat{d}_\pi\mathbf{e}_{\pi}+\hat{d}_+\mathbf{e}_{+}^*+\hat{d}_-\mathbf{e}_{-}^*\label{eq:d}
\end{align}
with
\begin{align}
    \hat{d}_\pi & = \hat{d}_z\label{eq:d_pi}\\
    \hat{d}_+   & = -\frac{1}{\sqrt{2}}\left(\hat{d}_x+i\hat{d}_y\right)\label{eq:d_+}\\
    \hat{d}_-   & = \frac{1}{\sqrt{2}}\left(\hat{d}_x-i\hat{d}_y\right)\label{eq:d_-}
\end{align}
and
\begin{align}
    \mathbf{e}_{\pi} & = \mathbf{e}_{z}\nonumber\\
                     & = \mathbf{e}_{3}\cos\iota+\mathbf{e}_{\Vert}\sin\iota\label{eq:e_pi}\\
    \mathbf{e}_{+} & = -\frac{1}{\sqrt{2}}\left(\mathbf{e}_x+i\mathbf{e}_y\right)\nonumber\\
                   & = \frac{1}{\sqrt{2}}\left(\mathbf{e}_3\sin\iota-\mathbf{e}_{\Vert}\cos\iota-i\mathbf{e}_{\bot}\right)\label{eq:e_+}\\
    \mathbf{e}_{-} & = \frac{1}{\sqrt{2}}\left(\mathbf{e}_x-i\mathbf{e}_y\right)\nonumber\\
                   & = \frac{1}{\sqrt{2}}\left(-\mathbf{e}_3\sin\iota+\mathbf{e}_{\Vert}\cos\iota-i\mathbf{e}_{\bot}\right),\label{eq:e_-}
\end{align}
where the different unit vectors on the right-hand side are defined in Figure \ref{fig:ARS_coordinates}. It is therefore advantageous to use the orientation of the magnetic field projected on the plane of the sky to define our polarisation basis. 

As was mentioned in Sec. \ref{sec:ARS}, given that we expect the polarisation basis to tend to a purely linear polarisation basis when $\iota\rightarrow\pi/2$ and a circular basis when $\iota\rightarrow0$ or $\pi$ we write
\begin{align}
    \mathbf{e}_1 & = a\mathbf{e}_\Vert+i b\mathbf{e}_\bot\label{eq:ell_bas1}\\
    \mathbf{e}_2 & = -b\mathbf{e}_\Vert+i a\mathbf{e}_\bot,\label{eq:ell_bas2}
\end{align}
where $a$ and $b$ are real functions of $\iota$ while $a^2+b^2=1$. The expressions central to our analysis, and which stems from the interaction Hamiltonian, concerns the relative phase shift accrued between these two polarisation states as the system evolves. More precisely, we are interested in the following relations
\begin{align}
    x_\pi\left(\iota\right) & = \left|\mathbf{e}_\pi\cdot\mathbf{e}_1\right|^2-\left|\mathbf{e}_\pi\cdot\mathbf{e}_2\right|^2\label{eq:x_pi}\\
    x_\pm\left(\iota\right) & = \left|\mathbf{e}_\pm^*\cdot\mathbf{e}_1\right|^2-\left|\mathbf{e}_\pm^*\cdot\mathbf{e}_2\right|^{2},\label{eq:x_+-}
\end{align}
which are the only parts that are functions of $\iota$ in the equations for the relative phase shift (see equations (\ref{eq:phi_p_pi})-(\ref{eq:phi_p_pm}) and (\ref{eq:Dphi_p})). Inserting equations (\ref{eq:e_pi})-(\ref{eq:ell_bas2}) into equations (\ref{eq:x_pi})-(\ref{eq:x_+-}) we find
\begin{align}
    x_\pi\left(\iota\right) & = \left(a^2-b^2\right)\sin^2\iota\label{eq:x_pi_2}\\
    x_\pm\left(\iota\right) & = \pm 2ab\cos\iota-\frac{1}{2}\left(a^2-b^2\right)\sin^2\iota\label{eq:x_+-_2}
\end{align}
and, from equations (\ref{eq:phi_p_pi})-(\ref{eq:phi_p_pm}) and (\ref{eq:Dphi_p}),
\begin{align}
    \Delta\phi_{p,\mathrm{s}} & \propto \left(a^2-b^2\right)\sin^2\iota\label{eq:Dphi_p,s_app}\\
    \Delta\phi_{p,\mathrm{c}} & \propto 2ab\cos\iota.\label{eq:Dphi_p,c_app}
\end{align}

Equations (\ref{eq:Dphi_p,s_app})-(\ref{eq:Dphi_p,c_app}) should exhibit the same symmetry as the Hamiltonian of the system. For example, it can be readily verified that $\Delta\phi_{p,\mathrm{s}}$ and $\Delta\phi_{p,\mathrm{c}}$ are unaffected by any rotation about the magnetic field axis $\mathbf{e}_z$ (i.e., the $C_\infty$ rotation group with $\mathbf{e}_z$ as the symmetry axis; \citealt{Bunker1998}). There exists an infinite number of combinations for $a$ and $b$ that satisfy this requirement. 

However, because our polarisation basis $\mathbf{e}_1$ and $\mathbf{e}_2$ (and corresponding states $\Ket{n_{p,1}}$ and $\Ket{n_{p,2}}$) are tied to the orientation of the magnetic field on the plane of the sky, the relative phase shifts $\Delta\phi_{p,\mathrm{s}}$ and $\Delta\phi_{p,\mathrm{c}}$ must also be invariant under rotations and improper rotations about the $\mathbf{e}_3$ axis, the inversion of coordinates and a reflection across the plane of the sky (i.e., the $D_{\infty\mathrm{h}}$ point group with $\mathbf{e}_3$ as the symmetry axis; \citealt{Bunker1998}) since they leave the inclination angle $\iota$ either unchanged or transformed to $\pi-\iota$. It follows that since $\sin^2\iota$ is invariant under these operations then so must be $a^2-b^2$, which must therefore take the form $1,\sin\iota,\cos^2\iota,...$ or some combination of these. On the other hand, $\cos\iota$ changes sign under improper rotations (including the inversion) and a reflection across the plane of the sky, thus so must $ab$. Considering these constraints the simplest relations for our basis require $a\propto\pm\cos\iota$ and $b\propto\pm1$ (or vice-versa). The elliptical basis we choose
\begin{align}
    & \Ket{n_{p,1}} = \frac{-1}{\sqrt{1+\cos^2\iota}}\left(\cos\iota\Ket{n_{p,\Vert}}+i\Ket{n_{p,\bot}}\right)\label{eq:n_1_app} \\
    & \Ket{n_{p,2}} = \frac{1}{\sqrt{1+\cos^2\iota}}\left(\Ket{n_{p,\Vert}}-i\cos\iota\Ket{n_{p,\bot}}\right) \label{eq:n_2_app}.
\end{align}
possesses the proper limits when $\iota\rightarrow 0,\pi/2$ or $\pi$ (see equations (\ref{eq:n_R})-(\ref{eq:n_L})). 

\section{Stokes parameters}\label{app:stokes}

The normalized Stokes parameters at frequency $\omega$ introduced in equations (\ref{eq:Stokes_q})-(\ref{eq:Stokes_v}) for an arbitrary incident signal can be explicitly calculated as a function of the different parameters entering in the analysis. That is, by expressing the relative phase shift accrued between the two elliptical basis states $\Ket{n_{p,1}}$ and $\Ket{n_{p,2}}$ with its real and imaginary parts
\begin{equation}
    \Delta\phi_p = \Delta\eta_p+i\Delta\gamma_p.\label{eq:Delta_phi_C}
\end{equation}
In the most general case, our elliptical basis takes the form
\begin{align}
    \Ket{n_{p,1}} &= a\Ket{n_{\Vert}}+ib\Ket{n_{\bot}}\label{eq:n1_app}\\
    \Ket{n_{p,2}} &= -b\Ket{n_{\Vert}}+ia\Ket{n_{\bot}}\label{eq:n2_app}
\end{align}
with $a$ and $b$ some real numbers ($a^2+b^2=1$), while the incident signal is
\begin{align}
    \Ket{n_{p,i}} = \alpha\Ket{n_{\Vert}}+\beta\Ket{n_{\bot}}\label{eq:ni_app}
\end{align}
where $\alpha$ and $\beta$ are (potentially) complex numbers obeying $\left|\alpha\right|^2+\left|\beta\right|^2=1$. We can then write for the Stoke parameters
\scriptsize
\begin{align}
    q_p\cdot i_p & = \left|\Braket{n_{p,\Vert}|n_{p,f}}\right|^2-\left|\Braket{n_{p,\bot}|n_{p,f}}\right|^2\nonumber\\
    & = \left[\left|\alpha\right|^2\left(e^{-2\Delta\gamma_p}a^4+b^4\right)-\left|\beta\right|^2\left(a^4+e^{-2\Delta\gamma_p}b^4\right)\right]\nonumber\\
    & \quad -\left(\left|\alpha\right|^2-\left|\beta\right|^2\right)a^2b^2\left(1+e^{-2\Delta\gamma_p}-4e^{-\Delta\gamma_p}\cos\left(\Delta\eta_p\right)\right)\nonumber\\
    & \quad -4ab\,\mathrm{Im}\left\{\alpha^*\beta\left[\rule{0mm}{3mm}e^{-\Delta\gamma_p}e^{-i\Delta\eta_p}a^2-e^{-\Delta\gamma_p}e^{i\Delta\eta_p}b^2\right.\right.\nonumber\\
    & \left.\left.\qquad\qquad\qquad +\frac{1}{2}\left(b^2-a^2\right)\left(1+e^{-2\Delta\gamma_p}\right)\right]\right\}\label{eq:Stokes_q_C}\\
    u_p\cdot i_p & = \left|\Braket{n_{p,+45}|n_{p,f}}\right|^2-\left|\Braket{n_{p,-45}|n_{p,f}}\right|^2\nonumber\\
    & = 2\,\mathrm{Re}\left\{\alpha\beta^*\left[e^{-\Delta\gamma_p}e^{i\Delta\eta_p}a^4+e^{-\Delta\gamma_p}e^{-i\Delta\eta_p}b^4+a^2b^2\left(1+e^{-2\Delta\gamma_p}\right)\right]\right.\nonumber\\
    & \left.\quad\quad -\alpha^*\beta a^2b^2\left[1+e^{-2\Delta\gamma_p}-2e^{-\Delta\gamma_p}\cos\left(\Delta\eta_p\right)\right]\right\}\nonumber\\
    & \quad -2\left(\left|\alpha\right|^2-\left|\beta\right|^2\right)ab e^{-\Delta\gamma_p}\sin\left(\Delta\eta_p\right)\label{eq:Stokes_u_C}\\
    v_p\cdot i_p & = \left|\Braket{n_{p,R}|n_{p,f}}\right|^2-\left|\Braket{n_{p,L}|n_{p,f}}\right|^2\nonumber\\
    & = 2\,\mathrm{Re}\left\{i\alpha\beta^*\left[e^{-\Delta\gamma_p}e^{i\Delta\eta_p}a^4+e^{-\Delta\gamma_p}e^{-i\Delta\eta_p}b^4+a^2b^2\left(1+e^{-2\Delta\gamma_p}\right)\right]\right.\nonumber\\
    & \left.\quad\quad -i\alpha^*\beta a^2b^2\left[1+e^{-2\Delta\gamma_p}-2e^{-\Delta\gamma_p}\cos\left(\Delta\eta_p\right)\right]\right\}\nonumber\\
    & \quad -2ab\left[\left(\left|\alpha\right|^2-\left|\beta\right|^2\right)\left(a^2-b^2\right)e^{-\Delta\gamma_p}\cos\left(\Delta\eta_p\right)\right.\nonumber\\
    & \quad\quad\left.-\left|\alpha\right|^2\left(e^{-2\Delta\gamma_p}a^2-b^2\right)+\left|\beta\right|^2\left(a^2-e^{-2\Delta\gamma_p}b^2\right)\right]\label{eq:Stokes_v_C}
\end{align}
\normalsize
\noindent with the normalization factor
\scriptsize
\begin{align}
    i_p & = \left|\Braket{n_{p,1}|n_{p,f}}\right|^2+\left|\Braket{n_{p,2}|n_{p,f}}\right|^2\nonumber\\
    & = \left(a^2\left|\alpha\right|^2+b^2\left|\beta\right|^2\right)e^{-2\Delta\gamma_p}+\left(b^2\left|\alpha\right|^2+a^2\left|\beta\right|^2\right)+2ab\left(1-e^{-2\Delta\gamma_p}\right)\mathrm{Im}\left\{\alpha\beta^*\right\}.\label{eq:Stokes_i_C}    
\end{align}
\normalsize

For the system we are considering in the main text, where  
\begin{align}
    a & = \frac{-\cos\iota}{\sqrt{1+\cos^2\iota}}\label{eq:a}\\
    b & = \frac{-1}{\sqrt{1+\cos^2\iota}}\label{eq:b}
\end{align}
and
\begin{align}
    \alpha & = \cos\varphi\label{eq:alpha}\\
    \beta & = \sin\varphi,\label{eq:beta}
\end{align}
equations (\ref{eq:Stokes_q_C})-(\ref{eq:Stokes_i_C}) become
\scriptsize
\begin{align}
    q_p\cdot i_p & = \frac{1}{\left(1+\cos^2\iota\right)^2}\left\{\cos^2\varphi\left(1+e^{-2\Delta\gamma_p}\cos^4\iota\right)-\sin^2\varphi\left(e^{-2\Delta\gamma_p}+\cos^4\iota\right)\right.\nonumber\\
    & \quad\quad \left.-\cos\left(2\varphi\right)\cos^2\iota\left[1+e^{-2\Delta\gamma_p}-4 e^{-\Delta\gamma_p}\cos\left(\Delta\eta_p\right)\right]\right\}\nonumber\\
    & \quad +2\frac{\cos\iota}{1+\cos^2\iota}e^{-\Delta\gamma_p}\sin\left(2\varphi\right)\sin\left(\Delta\eta_p\right)\label{eq:Stokes_q_main}\\
    u_p\cdot i_p & = e^{-\Delta\gamma_p}\left[\sin\left(2\varphi\right)\cos\left(\Delta\eta_p\right)-\frac{2\cos\iota}{1+\cos^2\iota}\cos\left(2\varphi\right)\sin\left(\Delta\eta_p\right)\right]\label{Stokes_u_main}\\
    v_p\cdot i_p & = \frac{\sin^2\iota}{1+\cos^2\iota}e^{-\Delta\gamma_p}\sin\left(2\varphi\right)\sin\left(\Delta\eta_p\right)\nonumber\\
    &\quad +2\frac{\cos\iota}{\left(1+\cos^2\iota\right)^2}\left[e^{-\Delta\gamma_p}\sin^2\iota\cos\left(2\varphi\right)\cos\left(\Delta\eta_p\right)\right.\nonumber\\
    & \quad\quad \left.-\left(1-e^{-2\Delta\gamma_p}\cos^2\iota\right)\cos^2\varphi+\left(e^{-2\Delta\gamma_p}-\cos^2\iota\right)\sin^2\varphi\right]\label{eq:Stokes_v_main}
\end{align}
with
\begin{align}
       i_p & = \frac{1}{1+\cos^2\iota}\left[e^{-2\Delta\gamma_p}\left(\cos^2\varphi\cos^2\iota+\sin^2\varphi\right)+\left(\cos^2\varphi+\sin^2\varphi\cos^2\iota\right)\right].\label{eq:Stokes_i_main}
\end{align}
\normalsize

For an unpolarised signal equations (\ref{eq:Stokes_q_C})-(\ref{eq:Stokes_i_C}) simplify to 
\begin{align}
    \left<i_{p,1}\right> & = e^{-\Delta\gamma_p}\cosh\left(\Delta\gamma_p\right)\label{eq:i_unp_C}\\
    \left<q_{p,1}\right>\left<i_{p,1}\right> & =  \frac{\sin^2\iota}{1+\cos^2\iota}e^{-\Delta\gamma_p}\sinh\left(\Delta\gamma_p\right)\label{eq:q_unp_C}\\
    \left<u_{p,1}\right>\left<i_{p,1}\right> & = 0\label{eq:u_unp_C}\\
    \left<v_{p,1}\right>\left<i_{p,1}\right> & = \frac{-2\cos\iota}{1+\cos^2\iota}e^{-\Delta\gamma_p}\sinh\left(\Delta\gamma_p\right)\label{eq:v_unp_C}
\end{align}
\noindent since $\left<\alpha\right>=\left<\beta\right>=\left<\alpha^*\beta\right>=0$ and $\left<\left|\alpha\right|^2\right>=\left<\left|\beta\right|^2\right>=1/2$ in this case. We have used the subscript ``$1$'' to designate the unpolarised nature of the signal, in accordance with the discussion in Sec. \ref{sec:generation}. It can be readily verified that equations (\ref{eq:i_unp_C})-(\ref{eq:v_unp_C}) lead to equations (\ref{eq:qp_unp})-(\ref{eq:vp_unp}). 

If a signal is composed of both a linearly polarised and an unpolarised components of respective intensity $I_{p,0}$ and $I_{p,1}$, then the incident state is, from Sec. \ref{sec:generation}, 
\begin{equation}
    \Ket{n_{p,i}} = \left(\alpha_0 + \alpha_1\right)\Ket{n_{p,\Vert}}+\left(\beta_0 + \beta_1\right)\Ket{n_{p,\bot}},\label{eq:pol_unpol_C}
\end{equation}
\noindent where 
\begin{align}
    & \alpha_i = \sqrt{\frac{I_{p,i}}{I_{p,0}+I_{p,1}}}\cos\left(\varphi_i\right)\label{eq:alpha_i_C}\\
    & \beta_i = \sqrt{\frac{I_{p,i}}{I_{p,0}+I_{p,1}}}\sin\left(\varphi_i\right)\label{eq:beta_i_C}
\end{align}
\noindent and $i=0,1$ for the polarised and unpolarised components, respectively. It follows that the total Stokes parameters, i.e., for the corresponding intensities of the compound signal, are given by
\begin{align}
    I_p = & I_{p,0}\,i_{p,0}+I_{p,1}\left<i_{p,1}\right>\label{eq:i_p_total_C}\\
    X_p = & I_{p,0}\,x_{p,0}\,i_{p,0}+I_{p,1}\left<x_{p,1}\right>\left<i_{p,1}\right>,\label{eq:x_p_total_C}
\end{align}

\noindent with $X_p=Q_p,U_p,V_p$ and similarly for $x_p$, and where $i_{p,0}$, $q_{p,0}$, etc., are calculated using equations (\ref{eq:Stokes_q_main})-(\ref{eq:Stokes_i_main}) and $\left<i_{p,1}\right>$, $\left<q_{p,1}\right>$, etc., with equations (\ref{eq:i_unp_C})-(\ref{eq:v_unp_C}).

\label{lastpage}

\end{document}